\documentclass[aps,prx,twocolumn,longbibliography,superscriptaddress,nofootinbib,10pt]{revtex4-2}

\usepackage{graphicx}
\usepackage{dcolumn}
\usepackage[english]{babel}
\usepackage{lipsum} 
\usepackage{times}
\usepackage{amsfonts,amssymb,bm}
\usepackage{amsmath}
\usepackage{epsfig}
\usepackage{mathrsfs}
\usepackage{color,soul}
\usepackage{bbold}
\usepackage{pstricks}
\usepackage{multirow}
\usepackage{textcomp}
\usepackage{braket}
\usepackage{verbatim}
\usepackage{empheq}
\usepackage[normalem]{ulem}
\definecolor{color1bg}{HTML}{b890a1}
\usepackage{xcolor}
\usepackage{rotating}
\usepackage{amsthm}
\usepackage{eurosym}
\usepackage{siunitx}
\usepackage{booktabs}
\usepackage[babel]{csquotes}

\usepackage{mwe}

\usepackage[framemethod=default]{mdframed}
\usepackage[font=footnotesize,labelfont=bf,labelsep=quad,justification=raggedright]{caption}
\usepackage{newfloat}

\DeclareFloatingEnvironment[name={S}]{suppfigure}
\DeclareFloatingEnvironment[name={S}]{supptable}
\allowdisplaybreaks
\usepackage{titlecaps}
\usepackage{placeins}

\newmdenv[linecolor=white,backgroundcolor=gray!15!]{myframe}

\definecolor{green1}{rgb}{0.33, 0.7, 0.69}

\usepackage[version=3]{mhchem}  
\usepackage{physics}
\usepackage{mathtools}
\usepackage{subcaption}
\usepackage{enumitem}

\newcommand{\figref}[1]{Fig.~\ref{#1}}

\usepackage{listings}
\usepackage{hyperref}
\hypersetup{
 colorlinks,
 citecolor=color1bg,
 filecolor=black,
 linkcolor=color1bg,
 urlcolor=black}

\begin{document}

\title{Interpretable rule-based learning in an autonomous thermodynamic network}

\author{Sparrow Suderman}
\thanks{sparrow.suderman03@gmail.com}
\affiliation{H. H. Wills Physics Laboratory, University of Bristol, Tyndall Avenue, Bristol, BS8 1TL, United Kingdom}

\author{Giulia Rubino}
\thanks{giulia.rubino@bristol.ac.uk}
\affiliation{H. H. Wills Physics Laboratory, University of Bristol, Tyndall Avenue, Bristol, BS8 1TL, United Kingdom}
\affiliation{Quantum Engineering Technology Labs, H. H. Wills Physics Laboratory and School of Electrical, Electronic, and Mechanical Engineering, University of Bristol, BS8 1FD, UK}

\date{\today}

\begin{abstract}
Machine learning is typically described in terms of deterministic logical operations, whereas physical systems generally operate in the presence of noise, dissipation and irreversibility. Here, we turn these physical effects into computational resources for an autonomous, interpretable learning architecture. We develop a classifier based on thermodynamic neurons, which are autonomous quantum thermal machines that implement logical operations through heat flow, and use these to construct a stochastic version of the Tsetlin machine, an interpretable rule-based learning architecture. By combining thermodynamic AND, NOT and OR gates with an autonomous coupling mechanism, we realise a learning system whose computation unfolds without the need for external time-dependent control. Despite its noisy components, the resulting classifier achieves classification accuracy that is statistically comparable to that of the standard Tsetlin machine. Reliability arises from architectural mechanisms such as thresholding and redundancy, rather than exact logical operations. Our results highlight that accurate and interpretable learning can emerge from autonomous stochastic dynamics, and establish thermodynamic computation as a viable framework for physical machine learning.
\end{abstract}

\maketitle

\section*{Introduction}

In conventional machine learning, computation is typically abstracted away from its physical implementation. Arithmetic operations are assumed to be exact, logic gates are considered reliable, and randomness is generally attributed to the data rather than the hardware performing the computation. This abstraction is remarkably effective for digital hardware, where the underlying physical implementation can often be neglected. Nevertheless, it also obscures the fact that information processing is ultimately a physical process, and is therefore subject to thermodynamic constraints~\cite{bennett1982thermodynamics,parrondo2015thermodynamics,seifert2012stochastic,wolpert2019stochastic,Tanaka2019}. This abstraction becomes less natural in emerging computing platforms such as thermodynamic, neuromorphic and quantum devices~\cite{nanophotonic_ML,echostates_ML,brain_ML}. In these systems, information processing is inherently noisy: switching events may be probabilistic, device states may fluctuate due to thermal or quantum effects, and variability can be part of the computational mechanism itself~\cite{Shim2017,deLeon2021,Woo2022}. Computation must then be understood as a process taking place within a noisy physical substrate, rather than as an ideal deterministic procedure later corrupted by noise. This perspective connects to broader efforts to understand the thermodynamic cost of computation~\cite{Deffner2013,Boyd2018,Faist2018}, to develop thermodynamically consistent models of information processing~\cite{seifert2012stochastic}, and to explore thermodynamic computing as an alternative paradigm for computation and AI~\cite{Conte2023,coles2023thermodynamic,aifer2024error,duffield2023thermodynamic}.

The central challenge, in this setting, shifts from suppressing noise in individual operations to ensuring that reliable computation and learning can emerge from inherently stochastic dynamics. In such regimes, it becomes important to understand how architectural principles such as redundancy, integration, thresholding and irreversibility can manage the trade-off between energetic efficiency and computational reliability. In the present work, we address this question by combining two main ingredients. The first is the thermodynamic neuron introduced in Ref.~\cite{thermodynamic_neuron}, a quantum thermal machine that operates autonomously, in the sense that, once the thermal inputs are fixed, it relaxes without externally timed control and produces an output that can be interpreted as the result of a logical gate. In this model, the quantum description provides a microscopic account of the stochastic logic primitives, whereas the computation performed by the network is classical and Boolean. The second is the Tsetlin machine~\cite{intro_to_tsetlin,granmo2018tsetlin,berge2018interpretable_tm}, a rule-based learning architecture that operates on Boolean data and produces explicit logical clauses for classification. This framework has shown promising results across a range of applications, including healthcare, regression and pattern recognition~\cite{medical_tsetlin,TM-patternrecognition,regression_TM}. It is also particularly well suited to our setting, as its discrete logical structure is naturally compatible with thermodynamic gates, while its interpretability, low-memory usage and energy-efficient character make it attractive for physical implementations~\cite{TM-energy,Berge2019,Saha2022}.

In this work, we combine these two ingredients to construct what we call a \emph{thermodynamic rule-evaluation engine}: an autonomous network of thermodynamic AND, NOT and OR gates that implements the rule-evaluation stage of the Tsetlin machine. Its operation is stochastic, since both the individual gates and the propagation of signals between them are governed by probabilistic thermal transitions. Learning is therefore not carried out through a sequence of externally timed operations, but unfolds through noisy physical transitions within the network. Information is propagated between neurons by a probabilistic coupling mechanism, so that stochasticity enters directly at the level of computation. The resulting architecture therefore provides a concrete setting within which we study learning on a genuinely noisy physical substrate.

The central challenge in such a network is reliability. Because the communication of outputs between successive neurons is mediated by the excitation of a signal qubit coupled to a finite output bath, the transmission of logical values is inherently probabilistic. Individual gates can therefore fail with non-negligible probability, and such errors may accumulate across the network. In this work, we show that this challenge can nevertheless be overcome through simple architectural features. In particular, by introducing a finite thermalisation interval and a redundancy scheme at the level of the gates, stochastic errors can be strongly suppressed without eliminating noise at the level of the individual components.

Across a range of benchmark datasets, we find that the resulting thermodynamic implementation achieves classification accuracy comparable, within statistical uncertainty, to that of the standard Tsetlin machine, with modest improvements in mean accuracy for some datasets. This shows that accurate and interpretable learning need not rely on deterministic logic or externally imposed control. Instead, it can emerge from an autonomous architecture whose operation is governed by stochastic, dissipative dynamics.

The remainder of this paper is organised as follows. In Sec.~\ref{sec:background}, we introduce the Tsetlin machine and the thermodynamic neuron as the two building blocks of our framework. In Sec.~\ref{subsec:AutonomyTN}, we analyse the irreversible operation of an individual thermodynamic neuron through its entropy production. We then develop the coupling mechanism for the autonomous network (Sec.~\ref{subsec:coupling}), analyse the role of redundancy in ensuring reliable operation (Sec.~\ref{subsec:redundancy}), and show how the resulting network carries out rule evaluation for the Tsetlin machine (Sec.~\ref{subsec:Rule-Evaluation_Engine}) and its resource requirements (Sec.~\ref{subsec:res_requirements}). Finally, in Sec.~\ref{subsec:performance} we evaluate the classification performance of the thermodynamic implementation across a range of datasets, before turning to a discussion of the broader implications of this work.

%
%

\section{Framework}
\label{sec:background}

\subsection{Tsetlin-machine rule evaluation}
\label{subsec:Tsetlin_machine}

First introduced by M. L. Tsetlin in the 1960s~\cite{intro_to_tsetlin}, the Tsetlin machine is a method for learning logical patterns from Boolean data and using them for classification tasks. For example, given a dataset describing animals, the task may be to distinguish cats from birds based on properties such as having fur or being able to fly. The aim is then to learn patterns such that a given input can be correctly assigned to a set category. 

In a Tsetlin machine, each data point (e.g., each row in a dataset) is described by \textit{True}/\textit{False} properties called \emph{features} (for example, whether an animal has fur). For each feature, both the feature and its negation are considered (in our example, \emph{Fur} and $\neg$ \emph{Fur}, where $\neg$ denotes NOT), and these are referred to as \textit{literals}. Classification is then performed using logical `\textit{rules}', which take the form of $\land$ (AND) combinations of literals (for example, \emph{Has Fur} $\land \neg$  \emph{Can Fly} $\land \neg$ \emph{Lays Eggs}). Each literal is assigned an internal memory state, conventionally divided into two regions corresponding to the actions \emph{exclude} and \emph{include}. For example, one may represent the memory by states $1,\ldots,2N$, with states $1,\ldots,N$ selecting exclusion of the corresponding literal from the rule, and states $N+1,\ldots,2N$ selecting inclusion. The precise number of states is not fixed by the model; rather, it controls how easily the automaton changes action under feedback. States close to the boundary between the two regions correspond to a less stable decision, while states deep within either region correspond to a more stable decision to exclude or include the literal\footnote{During rule evaluation, all states within a given region are treated identically; the distance from the boundary determines only how resistant the corresponding `include' or `exclude' decision is to subsequent feedback.}. A rule is therefore the conjunction of all included literals and evaluates to \textit{True} only when all of them are satisfied. For example, the rule \emph{Has Fur} $\land \neg$ \emph{Can Fly} $\land \neg$ \emph{Lays Eggs} would evaluate to \textit{True} for an animal that has fur, cannot fly, and does not lay eggs, and can therefore be used to identify a cat as opposed to a bird.

Having described how rules are formed, we now turn to how they are learned. In the standard parametrisation used here, we take $N=5$, so that each literal has ten possible memory states. The machine is initialised with all literals at state $N=5$, namely at the edge of the exclusion region, adjacent to the inclusion region. Learning then proceeds by iteratively updating these memory states based on observed data. During training, each input is provided together with its correct class label (e.g., cat or bird). The type of feedback applied is then determined by comparing this label with the evaluation of the rule on that input (i.e., whether the rule returns \textit{True} or \textit{False}). Below we describe the updating mechanism corresponding to each case.

When the label indicates that the input belongs to the target class (for example, when the input is a cat and the task is to identify cats), `\textit{Type I}' feedback is applied. This feedback has two cases, depending on whether the rule evaluates to \textit{True} or \textit{False}.
If the rule evaluates to \textit{True}, it has correctly identified the input. The feedback then reinforces the literals that supported this classification, so that similar inputs are more likely to be recognised in the future.
If instead the rule evaluates to \textit{False} (e.g., if the input is a cat but the rule fails to activate), this indicates that the rule does not capture the relevant features of the target class, and its memory states are therefore reduced, effectively erasing the rule.
Over time, some literals obtain high memory states (values close to 10), solidifying their presence in the rule. These literals correspond to the common traits that are \textit{True} for all objects in the target class. Likewise, literals which obtain memory states close to 1 correspond to traits that are unimportant (for example, for this classification task, \emph{Black}). Note, however, that this is not the same as a property being systematically absent from the target class: in that case, the corresponding \textit{negated} literal is included in the rule.

We now consider the case where the input does not belong to the target class (e.g., when the input is a bird but the rule is intended to identify cats), whereupon `\textit{Type II}' feedback is used. This feedback is applied only when the rule evaluates to \textit{True}, i.e., when an incorrect match occurs (for example, if the input is not a cat but the rule activates, indicating that it is too broad and should be refined). In this case, literals that are \textit{True} for the input are weakened, as they contributed to this incorrect activation. For instance, if the rule activates for a bird, literals such as \emph{Can Fly} are suppressed so that similar inputs are less likely to trigger the rule in the future. Conversely, literals that are \textit{False} for the input (such as $\neg$ \emph{Can Fly}) are reinforced, encouraging the rule to become more selective and better distinguish the target class. If instead the rule evaluates to \textit{False}, no feedback is applied, as the rule correctly rejects the input.

Overall, these feedback mechanisms update the memory states of the literals, reinforcing useful patterns and suppressing incorrect ones. `\textit{Type I}' feedback is applied stochastically, whereas `\textit{Type II}' feedback is typically deterministic. This stochasticity is meant to prevent the rules from overfitting to individual examples by limiting how aggressively literals are updated. Through repeated updates, the machine converges to a set of rules that capture the relevant patterns in the data. These rules can then be used to classify new inputs, while remaining interpretable as explicit logical rules. 

In the context of this work, the key step is the evaluation of a rule (i.e., determining whether it evaluates to \textit{True} or \textit{False} for a given input) as this directly determines which feedback is applied. We will therefore focus on implementing this evaluation step using a network of thermodynamic logic gates (which we will call a `\textit{rule-evaluation engine}'). In contrast, the feedback mechanism will be retained in its classical form. For extensions in which the feedback itself is stochastic, see Ref.~\cite{stochastic_TM}.

\subsection{Thermodynamic-neuron logic}
\label{subsec:neuron} 

Having described how logical rules are constructed and evaluated in the Tsetlin machine, we now introduce thermodynamic neurons as stochastic implementations of the individual logical operations required for rule evaluation, forming the basis of our \emph{thermodynamic rule-evaluation engine}.

The basic idea of temperature-based computation is to encode logical values not in discrete voltage levels or occupation states, but in thermodynamic biases. In the setting considered here, inputs are encoded in terms of inverse temperatures $\beta=(k_B T)^{-1}$, where $T$ is the temperature and $k_B$ is the Boltzmann constant, which we set to unity throughout. We associate $\beta_{\max}$ and $\beta_{\min}$ with the logical values \textit{True} and \textit{False}, respectively. A logical operation is then realised by coupling input and reference baths, treated as ideal reservoirs with effectively infinite heat capacity, to a thermal machine whose finite-capacity output bath evolves in response to the resulting heat currents. In this sense, the output of the computation is not a projective measurement on a qubit, but the thermodynamic bias established in the output reservoir.

The thermodynamic neuron, first introduced in Ref.~\cite{thermodynamic_neuron}, provides a concrete realisation of this idea. It is an autonomous quantum thermal machine that maps input temperatures to an output temperature in such a way that the resulting input-output relation approximates a logical gate. Computation is therefore realised through physical relaxation to a steady state: given input temperatures, the system evolves autonomously and settles to an output temperature representing the result of the computation.

\begin{figure}[bth]
\centering
\includegraphics[width=.85\columnwidth]{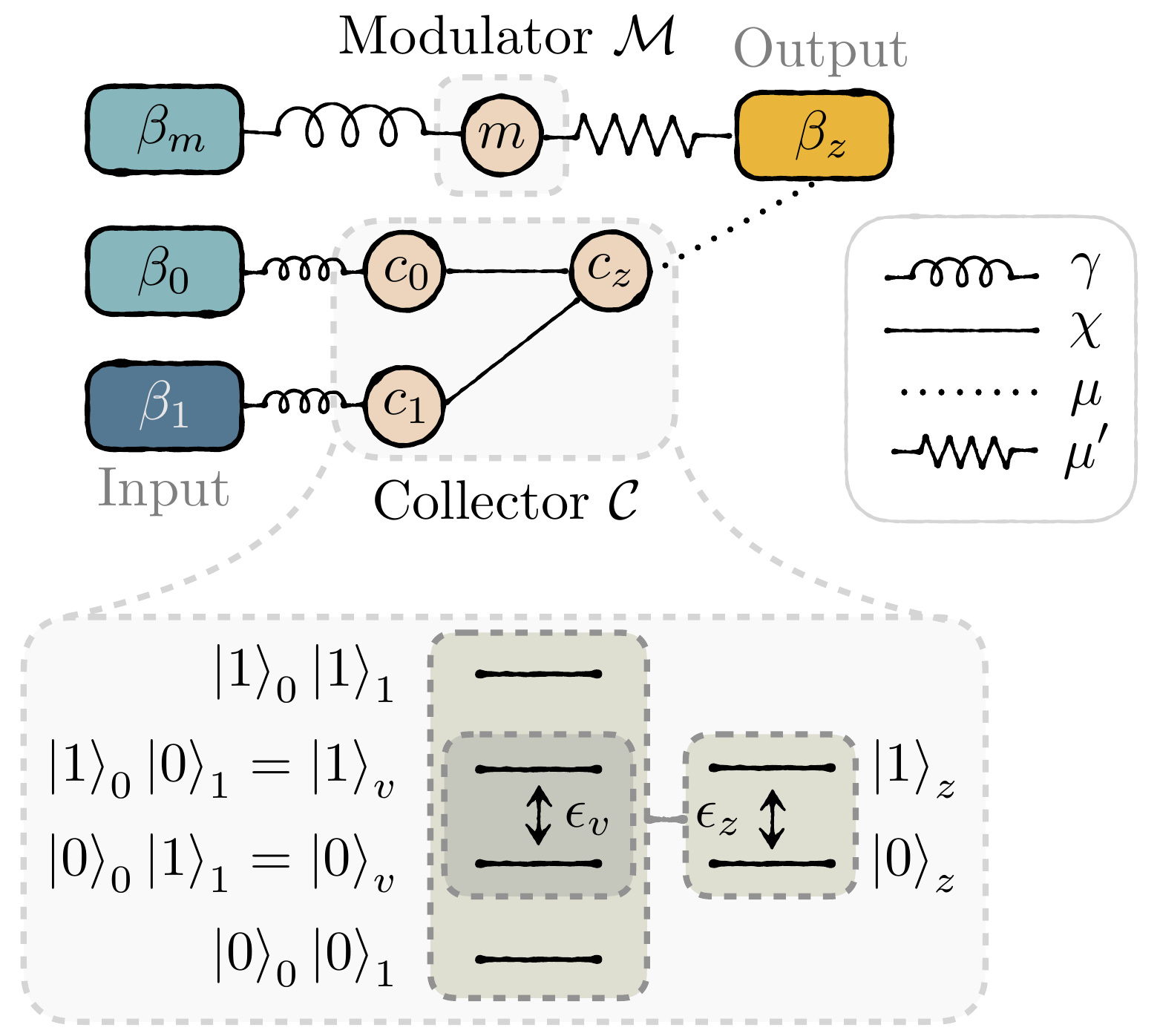}
\caption{\textbf{Thermodynamic neuron.} Autonomous quantum thermal machine composed of a collector and a modulator. The collector consists of three coupled qubits ($c_0, c_1, c_z$) interacting with thermal baths at inverse temperatures $\beta_0, \beta_1,$ and $\beta_z$, respectively. 
The states $\ket{0}_0\ket{1}_1$ and $\ket{1}_0\ket{0}_1$ define a virtual qubit with energy gap $\epsilon_v = \epsilon_z$, enabling resonant energy exchange with the output qubit $c_z$. 
The modulator qubit $m$, coupled to a bath at inverse temperature $\beta_m$, introduces the nonlinearity required for logical behaviour.}
\label{fig:thermodynamic_neuron}
\end{figure}

In order to implement such behaviour, the thermodynamic neuron combines two key components: a linear transformation of the inputs and a non-linear response that enables a decision between logical states. Indeed, the linear transformation alone would produce a continuous output, whereas logical operations require a separation into distinct states (e.g., \textit{True} or \textit{False}). The non-linear response achieves this by introducing a saturation effect, whereby a range of input values are mapped to the same extremal output, stabilising the system around well-defined logical states. These roles are played by two components of the machine, referred to as the \emph{collector} and the \emph{modulator} (\figref{fig:thermodynamic_neuron}).

The collector processes the input temperature and produces an effective `virtual' temperature to enable transformations that would not be possible using thermal baths alone. In fact, direct coupling to thermal reservoirs would restrict the output temperature to lie between the inputs, thereby limiting the range of achievable responses. By contrast, the collector can generate effective temperatures that lie outside this range, or even become negative, allowing for a much more flexible mapping from inputs to outputs. To realise this behaviour, each qubit is modelled as a two-level system with Hamiltonian $H_i = \epsilon_i \ket{1}\bra{1}$. When coupled to a thermal bath at inverse temperature $\beta_i$, each qubit relaxes to the Gibbs state $\tau(\beta_i) = e^{-\beta_i H_i}/\mathrm{Tr}\bigl(e^{-\beta_i H_i}\bigr)$, 
with excited-state population $g(\beta_i \epsilon_i) = (1 + e^{\beta_i \epsilon_i})^{-1}$.

Following Ref.~\cite{thermodynamic_neuron}, we illustrate the construction using the collector for a thermodynamic NOT gate, realised as a three-qubit machine~\cite{smallest_fridge}. Two qubits $c_0$ and $c_1$, with energy gaps $\epsilon_0 \geq \epsilon_1$, are coupled to thermal baths at inverse temperatures $\beta_0$ and $\beta_1$, respectively.
The joint system defines a \emph{virtual qubit} spanned by $\bigl\{\ket{0}_v = \ket{0}_0\ket{1}_1, \ket{1}_v = \ket{1}_0\ket{0}_1\bigr\}$,
with energy gap $\epsilon_v = \epsilon_0 - \epsilon_1$. The corresponding virtual temperature is defined through the population ratio $e^{-\beta_v \epsilon_v} = P(1_v)/P(0_v) = e^{-\beta_0 \epsilon_0}/e^{-\beta_1 \epsilon_1}$, yielding
\begin{equation}\label{eq:virtual_temp}
\beta_v = \frac{\beta_0 \epsilon_0 - \beta_1 \epsilon_1}{\epsilon_v}.
\end{equation}
The virtual temperature is thus a weighted combination of the input temperatures, but is not constrained to lie within their original range.

A third qubit $c_z$, with energy gap $\epsilon_z = \epsilon_v$, is coupled to a finite-capacity output bath $\mathcal{B}_z$, initially at inverse temperature $\beta_z$, and resonantly interacts with the virtual qubit through the energy-conserving Hamiltonian
\begin{equation}
H_{\mathrm{int}} = \chi\ket{1}\bra{0}_v\otimes\ket{0}\bra{1}_{z} + \mathrm{H.c.},
\end{equation}
where $\chi$ denotes the coupling strength~\cite{thermodynamic_neuron}. This interaction exchanges an excitation between the virtual qubit and $c_z$, coupling the degenerate states $\ket{1}_v\ket{0}_z$ and $\ket{0}_v\ket{1}_z$.
As the virtual qubit is effectively stabilised by its coupling to the thermal baths, it acts as a temperature source and drives the temperature of the output bath towards $\beta_v$. The resulting mapping exhibits an inversion behaviour, whereby increasing $\beta_1$ leads to a decrease in $\beta_z$.

The collector maps the input temperatures to the virtual temperature through a linear combination [Eq.~\eqref{eq:virtual_temp}]. The non-linear response required for logical behaviour is then introduced by the \textit{modulator}, which connects the output system to an additional qubit $m$, itself in contact with a reservoir at inverse temperature $\beta_m$.
To understand the effect of the modulator, it is useful to consider the heat flows acting on the output bath $\mathcal{B}_z$. The output temperature is determined by two competing contributions: one arising from the collector, mediated by $c_z$, which tends to drive the system towards the virtual temperature $\beta_v$, and one from the modulator, mediated by $m$, which pulls it towards $\beta_m$. The evolution of the output temperature is therefore governed by the balance between these competing heat currents.
To describe this quantitatively, we follow Ref.~\cite{thermodynamic_neuron} and model the interaction of each qubit with its thermal bath using a reset model, in which qubits undergo stochastic relaxation events that drive them towards thermal equilibrium. In this framework, an interaction with a bath at inverse temperature $\beta_k$ is described by the dissipator
\begin{equation}
\mathcal{L}^{(k)}(\rho) = \gamma_k \Bigl[ \mathrm{Tr}_k(\rho) \otimes \tau(\beta_k) - \rho \Bigr],
\end{equation}
where $\rho$ is the density operator of the full system, $\tau(\beta_k)$ is the Gibbs state, and $\gamma_k$ sets the rate of relaxation. The corresponding heat current is
\begin{equation}
j_k = \mathrm{Tr}\Bigl( H \mathcal{L}^{(k)}(\rho) \Bigr)
= \gamma_k \epsilon_k \bigl[ g(\beta_k \epsilon_k) - p_k \bigr],
\end{equation}
where $H$ is the Hamiltonian of the thermodynamic neuron, comprising the local qubit Hamiltonians and the interaction Hamiltonian, $H=\sum_{i \, \in \, {0,1,z,m}} H_i+H_{\mathrm{int}}$, and $p_k$ is the excitation probability of the \textit{k}-th qubit, so that the heat current is proportional to the difference between the actual population and the thermal population imposed by the bath\footnote{The qubit-bath interactions are modelled using the reset model, which is valid in the weak-coupling regime. In this regime, equilibration with the input and reference baths occurs more quickly than heat exchange with the finite output bath. Consequently, the collector and modulator remain at approximately the effective temperatures $\beta_v$ and $\beta_m$, while the temperature of the output bath $\beta_z(t)$ changes more slowly. This allows them to be treated as effective temperature sources for the output bath.}~\cite{thermodynamic_neuron}.

Applying this to the collector and modulator, which act as effective baths at inverse temperatures $\beta_v$ and $\beta_m$, respectively, and taking the output qubit to have instantaneous temperature $\beta_z(t)$ yield the heat currents $j_\mathcal{C}$ and $j_\mathcal{M}$ associated with the collector and modulator, respectively,
\begin{subequations}
\begin{align}
j_\mathcal{C} &= \mu \epsilon_z \Bigl[ g_z\bigl(\beta_z(t)\bigr) - g_z(\beta_v) \Bigr], \label{eq:heatCurrent_C}\\
j_\mathcal{M} &= \mu' \epsilon_z \Bigl[ g_z\bigl(\beta_z(t)\bigr) - g_z(\beta_m) \Bigr],
\label{eq:heatCurrent_M}
\end{align}
\end{subequations}
where $g_z(x) = g(\epsilon_z x)$. Here, $\mu$ is the effective coupling rate between the collector output qubit $c_z$ and the finite output bath $\mathcal{B}_z$, while $\mu'$ is the corresponding coupling rate between the modulator qubit $m$ and $\mathcal{B}_z$. These rates play the same role as the bath-coupling rates $\gamma_k$ introduced above, but refer specifically to the two channels through which heat is exchanged with the output bath.

The evolution of the output temperature then follows from the calorimetric relation, which links the net heat current to the rate of change of temperature via the heat capacity $C = \frac{dE}{dT}$,
\begin{equation}
\dot{\beta_z}(t) = -\frac{\beta_z^2}{C} \left(j_\mathcal{C} + j_\mathcal{M}\right).
\end{equation}
At steady state, the condition $j_{\mathcal{C}} + j_{\mathcal{M}} = 0$ determines the asymptotic temperature $\beta_z^\infty = \lim_{t \to \infty} \beta_z(t)$. Using Eqs.~\eqref{eq:heatCurrent_C}--\eqref{eq:heatCurrent_M}, the steady-state population satisfies
\begin{equation}
g_z(\beta_z^\infty) = \frac{\mu\, g_z(\beta_v) + \mu'\, g_z(\beta_m)}{\mu + \mu'},
\label{eq:excited_output}
\end{equation}
i.e., it is a weighted average of the populations associated with the collector and modulator.

To enforce logical behaviour, boundary conditions are imposed such that extremal values of the virtual temperature map to the logical states,
\begin{equation}
    \label{eq:MaxMin_definition}
    g_z(\beta_{\min/\max}) = \lim_{\beta_v \to \mp \infty} g_z(\beta_z^\infty).
\end{equation}
Since $g_z(\beta_v) \xrightarrow[\beta_v \rightarrow -\infty]{} 1$, $g_z(\beta_v) \xrightarrow[\beta_v \rightarrow +\infty]{} 0$, the steady-state population can be parameterised as
\begin{equation}
    g_z(\beta_z^\infty) = g_z(\beta_\text{max}) + \Delta g_z(\beta_v) \label{eq:g_z-beta_z-infty}
\end{equation}
with $\Delta:= g_z(\beta_\text{min}) - g_z(\beta_\text{max}) \geq 0$ (further details on this parametrisation are given in Supplementary~Note~\ref{SM:steady_state_out}). At the same time $g_z(\beta_z^\infty) = (1+\mathrm{e}^{\epsilon_z \beta_z^\infty})^{-1}$, thus
\begin{equation}\label{eq:beta_z^inf}
\beta_z^\infty = \frac{1}{\epsilon_z}\log\left[\frac{1}{g_z(\beta_{\max}) + g_z(\beta_v)\Delta} - 1\right].
\end{equation}
This expression can be approximated by a sigmoid function,
\begin{equation}
\beta_z^\infty \approx (1 + e^{-x})^{-1},
\end{equation}
with $x = (\epsilon_1 + \epsilon_z)(\beta_0 - \beta_1)$. The steepness of the sigmoid can be tuned via $\epsilon_z$, allowing control over the sharpness of the logical transition.
This non-linear behaviour produces a sharp transition between logical states, enabling the thermodynamic neuron to approximate a NOT gate.

The collector thus provides a flexible transformation of the inputs through the virtual temperature, while the modulator introduces the non-linearity required for logical behaviour. Together, these components realise a physical implementation of a logical gate. This framework extends naturally to other logical operations (e.g., AND, OR), as discussed in Supplementary Note~\ref{SM:virtual_temp}, and forms the basis of the stochastic learning network considered in the following sections.

The discussion so far has introduced the two ingredients underlying our approach: the rule-evaluation stage of the Tsetlin machine and the thermodynamic neuron as a physical logic primitive. We now turn to the main contribution of this work, namely, their integration into an autonomous stochastic rule-evaluation engine.

\section{Results}

We now present the implementation of a stochastic network of thermodynamic neurons that realises the rule-evaluation dynamics underlying the Tsetlin machine learning algorithm. The resulting architecture is inspired by autonomous quantum engines and operates without external time-dependent control. In contrast to the standard Tsetlin machine, which relies on externally timed, discrete updates, the present model evolves through spontaneous, time-independent transitions governed by its internal thermodynamic structure. Computation thus emerges as a cascade of intrinsically stochastic events.

In order to produce directed computation, the dynamics must be biased in order to favour forward over reverse transitions, thereby establishing a well-defined output. In our model, this bias arises from nonequilibrium conditions generated by temperature gradients between reference baths and finite-capacity output baths. These gradients induce steady heat currents through the network, providing the free energy required to sustain directed information processing. At the thermodynamic level, this bias manifests as irreversibility, i.e., an imbalance between forward and reverse transitions that establishes a preferred direction of information flow. In the following, we quantify this behaviour by analysing the entropy production of a single thermodynamic neuron and then show how neurons can be coupled to realise an autonomous thermodynamic rule-evaluation engine.

\vfill

\subsection{Autonomy of thermodynamic neuron}
\label{subsec:AutonomyTN}

For the thermodynamic neuron, the total entropy production satisfies~\cite{thermodynamic_neuron} 
\begin{equation} 
\dot{\Sigma} = \dot{S}_{\mathrm{sys}} - \sum_{\alpha} \beta_\alpha j_\alpha(t) \geq 0, 
\label{eq:rate of entropy prod}
\end{equation} 
where $\dot{S}_{\mathrm{sys}} := -\mathrm{Tr}\bigl[\dot{\rho} \log \rho\bigr]$ is the rate of change of the von Neumann entropy of the system degrees of freedom, with $\rho$ denoting the density operator of the full set of qubits in the thermodynamic neuron, excluding the thermal baths. The second term represents the corresponding entropy flow into the environment, with $j_\alpha(t)$ denoting the heat current flowing from the system into the bath at inverse temperature $\beta_\alpha$.

In the steady-state regime, defined by $\dot{\rho}=0$ (or, equivalently, $\dot{S}_{\mathrm{sys}}=0$), the entropy production is entirely determined by the heat currents, $\dot{\Sigma} = \dot{\Sigma}_\mathcal{C} + \dot{\Sigma}_\mathcal{M}$, with
\begin{subequations}
    \begin{align}
    \label{eq:entropy_productionC}
     \dot{\Sigma}_\mathcal{C} &= -\beta_0j_0-\beta_1j_1-\beta_zj_\mathcal{C}, \\
     \dot{\Sigma}_\mathcal{M} &= -\beta_mj_m-\beta_zj_\mathcal{M}.
     \label{eq:entropy_productionM}
     \end{align}
\end{subequations}
To proceed, we express the entropy production in terms of the coarse-grained currents $j_\mathcal{C}$ and $j_\mathcal{M}$ and analyse the collector and modulator contributions separately.

For the collector, energy-preserving transitions of the form $\ket{1}_0 \ket{0}_1 \ket{0}_z \leftrightarrow \ket{0}_0 \ket{1}_1 \ket{1}_z$ imply that the loss of energy $\epsilon_0$ by qubit $c_0$ is accompanied by energy gains $\epsilon_1$ and $\epsilon_z$ in qubits $c_1$ and $c_z$, respectively. Energy conservation therefore enforces $\epsilon_0 = \epsilon_1 + \epsilon_z$, and the corresponding heat currents satisfy $j_0 + j_1 + j_\mathcal{C} = 0$. Since each transition exchanges fixed energy quanta, the heat currents are proportional, $\frac{j_0}{\epsilon_0} = - \frac{j_1}{\epsilon_1} = - \frac{j_\mathcal{C}}{\epsilon_z}$. The entropy production of the collector can thus be expressed as
\begin{equation}
    \dot{\Sigma}_\mathcal{C}  = \biggl[\frac{\epsilon_0}{\epsilon_z} \beta_0 -\frac{\epsilon_1}{\epsilon_z} \beta_1 - \beta_z \biggr]j_\mathcal{C} 
    = \bigl[\beta_v - \beta_z \bigr] j_\mathcal{C}
\end{equation}
where we use Eq.~\eqref{eq:virtual_temp} and the fact that $\epsilon_v = \epsilon_0 - \epsilon_1 = \epsilon_z$. Using the steady-state solution for $\beta_z$ derived in Sec.~\ref{subsec:neuron} [Eq.~\eqref{eq:beta_z^inf}], and noting that $g_z(\beta)$ is monotonically decreasing, it follows that $j_\mathcal{C}$ has the same sign as $\beta_v - \beta_z$ and, consequently, $\dot{\Sigma}_\mathcal{C} \geq 0$.

For the modulator, only two baths are involved, so at steady state $j_m + j_\mathcal{M} = 0$. Therefore,
\begin{equation}
    \dot{\Sigma}_\mathcal{M} = (\beta_m - \beta_z) j_\mathcal{M}.
\end{equation}
Since $g_z(\beta)$ is monotonically decreasing, $j_\mathcal{M}$ has the same sign as $\beta_m - \beta_z$, and therefore $\dot{\Sigma}_\mathcal{M} \geq 0$.

This demonstrates the irreversible operation of an individual thermodynamic neuron.

\subsection{Coupling mechanism}
\label{subsec:coupling}

We now turn to the coupling mechanism underlying the autonomous network of thermodynamic neurons. This is realised by introducing a signal qubit subject to first-passage-time dynamics under continuous monitoring~\cite{Gardiner1985,Gillespie1991,Garrahan2017,Ptaszynski2018,Kewming2024} (\ref{subsec:signal}), together with an autonomous quantum clock that generates timing signals through its own thermodynamically driven dynamics and thereby enforces finite thermalisation times (\ref{subsec:clock}). We further show how redundancy can be used to enhance the reliability of logical operations, and evaluate the performance of the resulting network across a range of datasets, demonstrating that the stochastic implementation achieves accuracy comparable to that of the standard Tsetlin machine.

\begin{figure}[bth]
    \centering
    \includegraphics[width=\columnwidth]{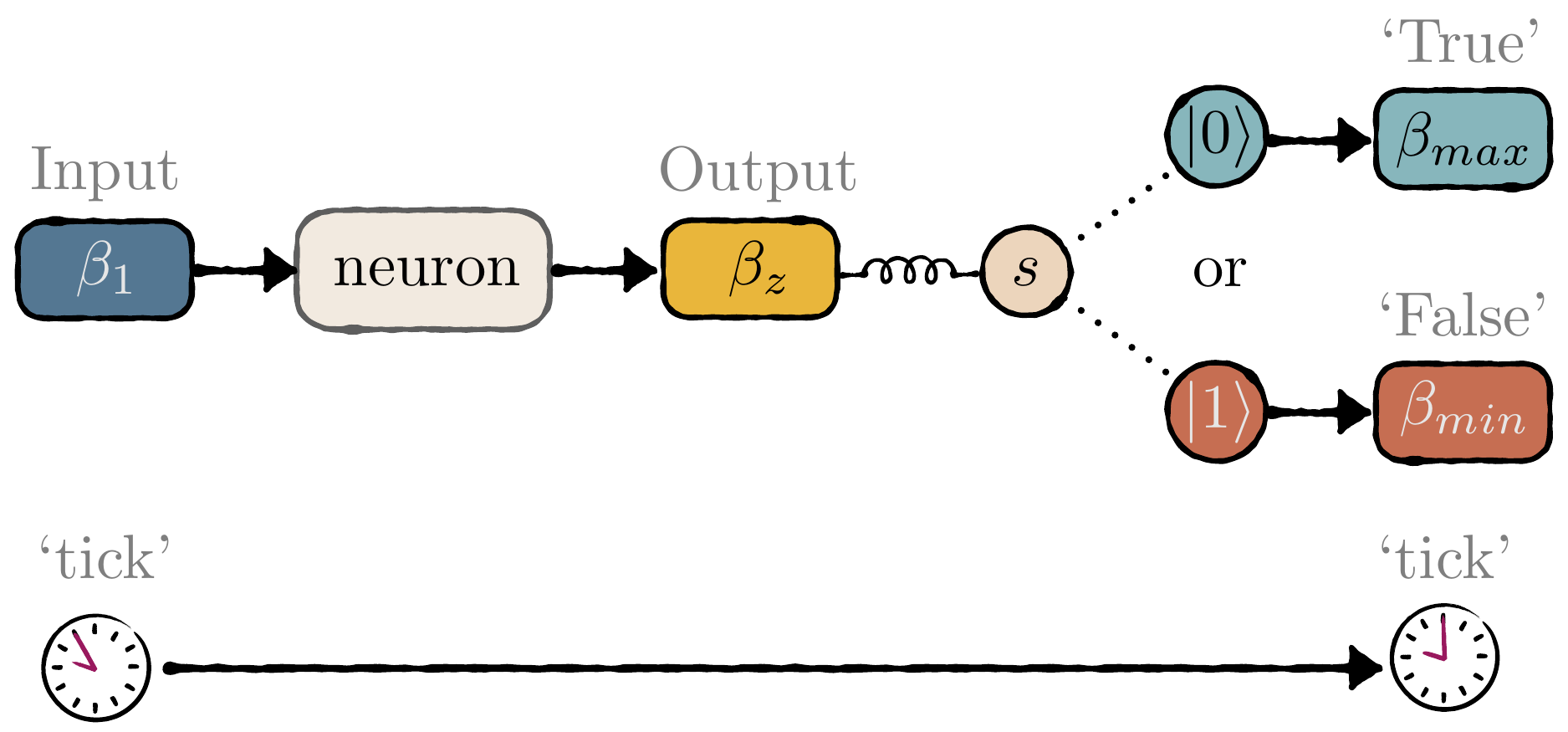}
    \caption{\textbf{Coupling mechanism for thermodynamic neurons.} A finite thermalisation interval defines the observation window for the neuron. During this time, a signal qubit $s$, coupled to the output bath $\mathcal{B}_z$, undergoes stochastic excitation dynamics determined by the bath temperature. The neuron output is inferred from a first-passage event: excitation of the signal qubit within the interval yields \textit{False}, while its absence yields \textit{True}. This output is then used to set the input of the subsequent neuron, enabling autonomous signal propagation through the network.}
    \label{fig:coupling}
\end{figure}

\subsubsection{First-passage readout}
\label{subsec:signal}

To construct an autonomous network, the signalling mechanism linking successive neurons must itself operate without external control. However, because input baths are modelled as having infinite heat capacity while output baths are finite, a given bath cannot simultaneously serve as both input and output. A natural approach would be to measure the state of the output bath and externally set the subsequent input, but this would introduce external control and thus compromise autonomy.
We therefore introduce the coupling mechanism shown in Fig.~\ref{fig:coupling}, which preserves autonomy while enabling probabilistic signal propagation across the network.

A signal qubit $s$, with energy gap $\epsilon_s$, is coupled to the output bath $\mathcal{B}_z$ and continuously monitored over a finite thermalisation interval. The neuron output is then inferred from the occurrence of excitation events during this interval: if the signal qubit becomes excited at any time, the output is taken to be \textit{False}, while the absence of excitation corresponds to \textit{True}. In this way, the readout is implemented through a first-passage process.

\begin{figure}[hbt]
    \centering
    \includegraphics[width=\linewidth]{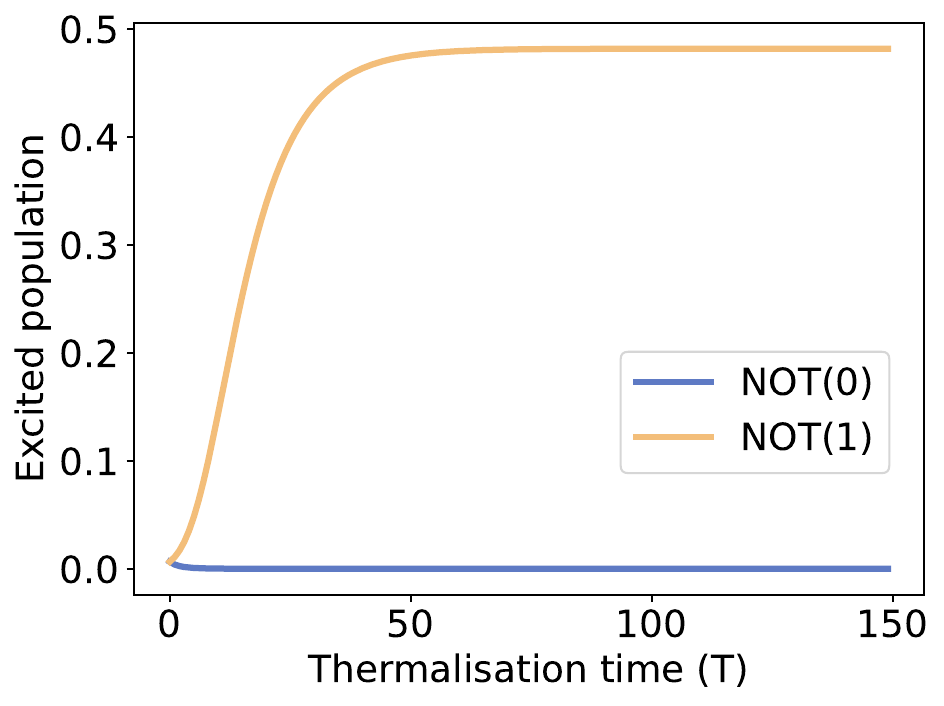}
    \caption{\textbf{Excited-state population of the signal qubit for a NOT gate.} The excitation probability $g\bigl(\beta_z(t)\epsilon_s\bigr)$ is shown for logical inputs \textit{False} (blue) and \textit{True} (orange). Lower values of $\beta_z(t)$ (corresponding to \textit{False}) lead to higher excitation probability, while larger values (\textit{True}) suppress excitations. Over a finite observation interval, this leads to a higher probability of observing at least one excitation event for \textit{False}, enabling a probabilistic readout of the neuron output. The curves are shown for $T = 10^6$ (in dimensionless relaxation-time units) and $\epsilon_s = 10$.}   
    \label{fig:excited_pop_NOT}
\end{figure}

During thermalisation, the excitation probability of the signal qubit depends on the instantaneous temperature of the output bath,
\begin{equation}
    g\bigl(\beta_z(t) \,\epsilon_s\bigr) = \frac{1}{1+\mathrm{e}^{\beta_z(t) \,\epsilon_s}},
    \label{eq:excitation prob}
\end{equation}
where the finite heat capacity of the output bath enters through the calorimetric evolution of $\beta_z(t)$. The signal qubit is therefore treated as a weak probe of this time-dependent bath: its excitation probability follows the instantaneous thermal value $g(\beta_z(t)\epsilon_s)$. In our model, we neglect the backaction of the signal qubit on $\beta_z(t)$, which is justified when the heat exchanged with the signal qubit is small compared with the energy scale associated with the finite output bath. This behaviour is illustrated in Fig.~\ref{fig:excited_pop_NOT} for the NOT gate.

We initialise the output bath at the midpoint of the logical temperatures, $\beta_z(0) = (\beta_{\min} + \beta_{\max})/2$, so that the subsequent evolution of $\beta_z(t)$ is unbiased towards either logical state. The dynamics then drive $\beta_z(t)$ towards one of the two logical values, $\beta_{\min}$ or $\beta_{\max}$, depending on the input. The direction of this evolution thus encodes the neuron output. Specifically, from Eq.~\eqref{eq:excitation prob}, the excitation probability satisfies $0 \le g\bigl(\beta_z(t)\epsilon_s\bigr) \leq 1/2$, with the upper bound attained as $\beta_z(t) \to 0$ and the lower bound as $\beta_z(t) \to \infty$. As $\beta_{\max}$ and $\beta_{\min}$ correspond to \textit{True} and \textit{False}, respectively, excitation events occur more frequently when the neuron outputs \textit{False}. Over a finite observation interval, this results in a higher likelihood of detecting an excitation, allowing the signal qubit to act as a probabilistic indicator of the neuron output.

Treating the signal qubit as an additional dissipative degree of freedom coupled to the output bath introduces a contribution $-\beta_z j_s$ to the total entropy production [Eqs.~\eqref{eq:entropy_productionC}--\eqref{eq:entropy_productionM}]. Since the signal qubit rapidly thermalises with the output bath, its excitation probability remains close to the instantaneous thermal value, $p_s \approx g(\beta_z \epsilon_s)$, and the associated heat current $j_s \propto g(\beta_z \epsilon_s) - p_s$ is therefore negligible. This is consistent with the weak-probe approximation introduced above. The entropy production thus remains non-negative, so that autonomy is preserved.

However, this construction does not by itself guarantee reliable output extraction. In particular, the excitation probability is bounded above by $1/2$, so that even for a \textit{False} output, excitation is not certain. As a result, the neuron exhibits an inherent bias towards outputting \textit{True}. 

This limitation is mitigated by exploiting the first-passage nature of the readout. Rather than relying on instantaneous excitation probability, the output is determined over a finite observation interval of duration $T$, and thus depends on the probability that an excitation occurs within this time window. The choice of $T$ (expressed in dimensionless relaxation-time units in our simulations) balances two competing effects: it must be sufficiently long to allow excitation when the correct output is \textit{False}, while remaining short enough to suppress unwanted excitation when the correct output is \textit{True}. In this way, $T$ acts as a control parameter that compensates for the intrinsic bias of the signal qubit.

In Tab.~\ref{tab:T_network}, we evaluate the probability of perfect network performance (i.e., no incorrect gate outputs) as a function of $T$. We find that this probability is maximised for $T = 50$, and therefore adopt this value for all subsequent simulations. Further details on the choice of thermalisation time, including its effect on individual gate reliabilities and network-level error accumulation, are provided in App.~\ref{app:thermalisationtime}.

\begin{table}[htb]
    \centering
        \begin{tabular}{c|cccc}
    \toprule
      T & 25 & \textbf{50} & 100 & 150  \\
      \midrule
    P(error free) & 0.9184 & \textbf{0.9259} & 0.9172 & 0.9087\\
    P(gate failure) & 0.08164 & \textbf{0.07409} & 0.08277 & 0.09132\\
    \bottomrule
    \end{tabular}
    \caption{\textbf{Probability of perfect computation and of at least one gate failure for the single-feature circuit (see Sec.~\ref{subsec:Rule-Evaluation_Engine} and Fig.~\ref{fig:network}).} 
    The probability of perfect computation (i.e., no gate failure) is obtained as the product of the individual gate success probabilities, reflecting the number of occurrences of each gate in the circuit: $\text{P(NOT correct)}^3 \cdot \text{P(OR correct)}^2 \cdot \text{P(AND correct)}$. Here, $T$ denotes the number of thermalisation steps, with each step corresponding to a fixed dimensionless thermalisation interval of $10^6$ relaxation-time units. The optimal value of $T$ is highlighted in bold.}
    \label{tab:T_network}
\end{table}

At this stage, however, the observation interval $T$ is introduced as an external parameter. To achieve a fully autonomous implementation, this timescale must instead arise from the intrinsic dynamics of the system. In the next section, we show how this can be accomplished by introducing an autonomous timing mechanism that replaces the externally imposed observation window.

\subsubsection{Clock-controlled thermalisation}
\label{subsec:clock}

To enforce a finite thermalisation time, a mechanism is required to define the duration $T$ of each neuron's evolution. Importantly, this must be achieved without external time-dependent control in order to preserve autonomy. This motivates the use of autonomous quantum clocks, which have been studied as physical timekeeping devices in thermodynamic settings~\cite{clock,Woods2019,Milburn2020,Pearson2021}. In the present work, we employ the autonomous quantum clock introduced in Ref.~\cite{clock}. The clock is realised by coupling hot and cold reservoirs, such that the resulting temperature gradient drives a steady current. This current excites a load, which is lifted along a ladder of energy levels until it reaches an unstable highest-energy state. Its subsequent decay is accompanied by the emission of a photon, which is registered as a `tick' of the clock.

Within the network, clock ticks define the thermalisation interval: the evolution of each neuron is terminated by the occurrence of a tick, thereby setting the observation window.
At the same time, the signal qubit undergoes stochastic excitation dynamics, with the first excitation event occurring at a random time $\tau_{\mathrm{exc}}$. The readout is thus governed by the competition between two stochastic processes: excitation of the signal qubit and emission of a clock tick. The relevant stopping time is given by $\tau = \min\{\tau_{\mathrm{exc}}, \tau_{\mathrm{tick}}\}$, where $\tau_{\mathrm{tick}}$ denotes the time of the next clock tick.

If $\tau_{\mathrm{exc}} < \tau_{\mathrm{tick}}$, an excitation is detected before the clock ticks, and the neuron outputs \textit{False}. Conversely, if $\tau_{\mathrm{exc}} > \tau_{\mathrm{tick}}$, the clock tick occurs first, terminating the thermalisation interval without excitation, and the output is assigned \textit{True}.

Both $\tau_{\mathrm{exc}}$ and $\tau_{\mathrm{tick}}$ arise from stochastic, time-independent dynamics of the combined system, comprising the signal qubit, detector, clock, and control degrees of freedom. The measurement is therefore fully autonomous: both detection of excitation and termination of the observation window are generated internally by competing stochastic processes.
More generally, detection events, clock ticks, and the switching of input baths are all realised as transitions within the enlarged system. The updating of inputs and the progression through successive neurons are therefore not externally applied operations, but emerge from the intrinsic dynamics. The full network is governed by a single time-independent generator, and both measurement outcomes and sequencing of computational steps arise entirely from internal processes, without external timing or control.
Computation then proceeds through thermally driven transitions within individual neurons, with local temperature gradients biasing the system towards specific logical states. These transitions can trigger excitation of signal qubits and clock ticks, which in turn regulate the progression of thermalisation across successive layers. In this way, global computation unfolds as a self-propagating, thermodynamically driven process.

\subsection{Bias suppression by redundancy}
\label{subsec:redundancy}

While the first-passage readout mechanism introduced in Sec.~\ref{subsec:signal} reduces the intrinsic bias of the signal qubit, it does not eliminate it entirely. Recall indeed that the \textit{False} output is associated with the occurrence of an excitation event within the observation interval, whereas the \textit{True} output corresponds to the absence of excitation. Since excitation is a stochastic process with probability bounded above by $1/2$, missed excitations are more likely than spurious ones, resulting in a residual bias towards the \textit{True} output.
This asymmetry is quantified in Tab.~\ref{tab:redundancy}, where for each gate the probability of incorrectly outputting \textit{True} exceeds that of incorrectly outputting \textit{False} by several orders of magnitude. Even small biases at the level of individual neurons can accumulate across the multi-layer structure of the \textit{thermodynamic rule-evaluation engine} (detailed in Sec.~\ref{subsec:Rule-Evaluation_Engine}), where the outputs of earlier gates serve as inputs to subsequent ones. This issue becomes particularly pronounced as errors propagate through successive layers and compound, degrading overall performance.

\begin{table}[hbt]
    \centering
    \small
    \setlength{\tabcolsep}{2pt}
        \begin{tabular}{c|ccc}
    \toprule
     P(output$\mid$expected)  & NOT & AND & OR\\
      \midrule
    P(0$\mid$1) & $0.02237$ & $0.02569$ &  $0.07290$\\
    P(1$\mid$ 0) & $1.262\times10^{-9}$ & $1.260\times10^{-7}$ & $4.307\times10^{-10}$\\
    \bottomrule
    \end{tabular}
    \caption{Probabilities of incorrect outputs for each logical value and gate, calculated using Eq.~\ref{eq:excitation prob}. Results are shown for $T = 50$, with probabilities given to four significant figures.}
    \label{tab:redundancy}
\end{table}

To address this imbalance, we introduce redundancy as an error-correction mechanism~\cite{VonNeumann1956,redundancy, shor_redundancy, fault-tolerant-redundancy}. 
In particular, each logical gate is implemented using multiple identical neurons receiving the same input, and the output is determined collectively. If any signal qubit undergoes an excitation event, the output is \textit{False}, whereas \textit{True} is assigned only in the absence of excitation across all neurons. This corresponds to a logical $\lor$ (OR) over excitation events, or equivalently a logical $\land$ over \textit{True} outputs.

This construction selectively suppresses errors in the \textit{True} output while retaining an autonomous first-passage readout. The $N$ signal qubits are monitored in parallel, and the first excitation among them registers a \textit{False} output. An incorrect \textit{True} output therefore occurs only if all duplicated neurons fail to excite during the readout interval. If the probability of this error for a single neuron is $p$, then for $N$ redundant neurons it scales as $p^N$, yielding exponential suppression. Redundancy thus compensates for the intrinsic bias of individual neurons. The choice of the redundancy factor $N$ must balance the resulting error suppression against the associated resource overhead. Its effect on the overall classification accuracy is analysed in Sec.~\ref{subsec:performance}.

\subsection{Network architecture}
\label{subsec:Rule-Evaluation_Engine}

\begin{figure*}[hbt]
\begin{lstlisting}[language=Python]
def evaluate_rule(observation, literals):
    rule_satisfied = True
    for i in range(len(observation)):  # for each feature

        # literal requires X_i = True but feature is False
        if literals[0, 2*i] == 1 and observation[j] == 0:
            rule_satisfied = False
            break

        # literal requires X_i = False but feature is True
        elif literals[0, 2*i + 1] == 1 and observation[j] == 1:
            rule_satisfied = False
            break

    return rule_satisfied
\end{lstlisting}
\caption{\textbf{Standard rule evaluation.} 
Algorithm for checking whether an input (`observation') satisfies a rule expressed as a conjunction of `literals'. Each feature $X_i$ is compared against its associated literals $L_{2i}$ and $L_{2i+1}$, which specify whether the feature is required to be \textit{True} or \textit{False}, respectively. The function returns \textit{False} as soon as a violation is detected (i.e., a required literal is not satisfied), and \textit{True} otherwise.}
\label{fig:code}
\end{figure*}

Having established an autonomous coupling mechanism and strategies to ensure reliable information flow, we now consider how thermodynamic neurons can be arranged to implement the rule-evaluation part of the Tsetlin machine algorithm. As outlined in Sec.~\ref{subsec:Tsetlin_machine}, this step determines whether a given input satisfies a conjunction of literals, i.e., whether the rule evaluates to \textit{True} or \textit{False}. During learning, this outcome determines the type of feedback applied to the Tsetlin machine, while during classification it contributes a vote towards the predicted class.

We begin with the standard implementation. Consider an input represented by binary features $\{X_i\}$. For each feature $X_i$, the rule may include either the literal $X_i$ or its negation $\neg X_i$, encoded by the binary variables $L_{2i}$ and $L_{2i+1}$, respectively. Thus, $L_{2i}=1$ means that the rule requires $X_i=1$, while $L_{2i+1}=1$ means that it requires $X_i=0$. If $L_{2i}=L_{2i+1}=0$, then the feature is unconstrained, meaning that its value does not affect whether the rule is satisfied. The rule is satisfied only if none of these requirements is violated. Equivalently, for each feature one checks whether the observed value of $X_i$ is compatible with the included literal(s), and the overall rule evaluates to \textit{True} only if this holds for every feature. This procedure is illustrated in Fig.~\ref{fig:code}.

\begin{figure}[hbt]
    \centering
    \includegraphics[width=.75\linewidth]{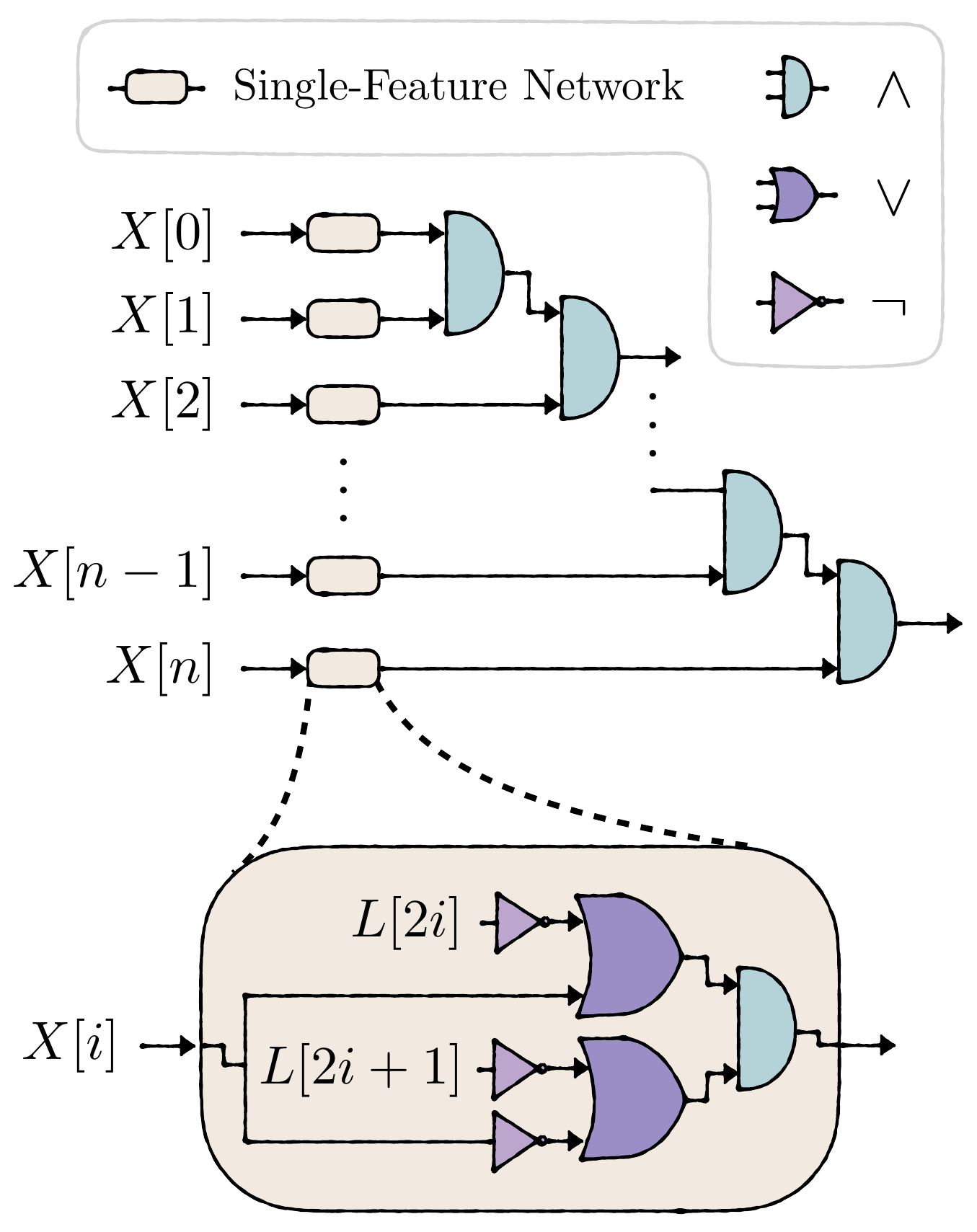}
    \caption{\textbf{Thermodynamic rule-evaluation engine.} 
    The network evaluates each feature $X_i$ independently using a single-feature network (rectangular boxes), and combines the results via a cascade of $\land$ gates. The overall output is \textit{True} if and only if all features satisfy the rule.  For each feature $X_i$, the single-feature network implements Eq.~\eqref{eq:logical_condition} (see Fig.~\ref{fig:code} for the standard implementation). Thus, the output is \textit{True} if and only if the input feature is consistent with all included literals.}
    \label{fig:network}
\end{figure}

In the thermodynamic setting, we reproduce the same logical operation physically using a network of thermodynamic neurons, forming our thermodynamic rule-evaluation engine (Fig.~\ref{fig:network}). The overall architecture mirrors the factorised structure of the standard algorithm. Each feature $X_i$ is first processed independently by a \emph{single-feature network}, which checks whether that feature is consistent with the literals associated with it. The outputs of all single-feature networks are then combined through a cascade of $\land$ gates. The final output is therefore \textit{True} if and only if every feature satisfies its corresponding literal constraints, exactly as in the standard rule evaluation.

The basic building block is the single-feature network. Its task is to evaluate, for a fixed feature $X_i$, whether the local consistency condition is satisfied. In logical form, this condition can be written as
\begin{equation}\label{eq:logical_condition}
(\neg L_{2i} \lor X_i)\land(\neg L_{2i+1}\lor \neg X_i).
\end{equation}
This expression has a simple interpretation. If $L_{2i}=1$, then the feature is required to be \textit{True}; if $L_{2i+1}=1$, it is required to be \textit{False}; and if both are zero, the feature is left unconstrained. The single-feature network outputs \textit{True} precisely when the input $X_i$ satisfies these requirements, and \textit{False} otherwise. Hence, it implements the same local check carried out in the standard algorithm, but in thermodynamic form.
For example, if $L_{2i}=1$ and $X_i=1$, the condition is satisfied and the output is \textit{True}; if instead $L_{2i}=1$ and $X_i=0$, the condition is violated and the output is \textit{False}. Similarly, if $L_{2i+1}=1$, the output is \textit{True} only when $X_i=0$. Finally, if neither literal is included, the feature contributes no constraint and the output is always \textit{True}. This behaviour is summarised in Table~\ref{tab:network results}.

\begin{table}[htb]
    \centering
    \begin{tabular}{ccc|c}
    \toprule
    \textbf{$X_i$} & \textbf{$L_{2i}$} & \textbf{$L_{2i+1}$} & \textbf{Output} \\
    \midrule
    0 & 0 & 0 & \textit{True} \\
    0 & 0 & 1 & \textit{True} \\
    0 & 1 & 0 & \textit{False} \\
    1 & 0 & 0 & \textit{True} \\
    1 & 0 & 1 & \textit{False} \\
    1 & 1 & 0 & \textit{True} \\
    \bottomrule
    \end{tabular}
    \caption{\textbf{Truth table for the single-feature network.}
    \(L_{2i}=1\) (\(L_{2i+1}=1\)) indicates that the feature is required to be \textit{True} (\textit{False}). If both are zero, the feature is unconstrained and always evaluates to \textit{True}  regardless of the value of $X_i$. The contradictory configuration $L_{2i}=L_{2i+1}=1$, in which both $X_i$ and $\neg X_i$ are included simultaneously, is omitted.}
    \label{tab:network results}
\end{table}

The full thermodynamic rule-evaluation engine is obtained by composing these single-feature consistency checks across all features and combining them with thermodynamic $\land$ gates. In this way, the network realises the same rule-evaluation logic as the standard Tsetlin machine, while the feedback and automaton updates remain classical.

\subsection{Resource requirements}
\label{subsec:res_requirements}

Before evaluating the network's performance, we examine the resource requirements of the thermodynamic rule-evaluation engine, focusing specifically on the number of infinite- and finite-capacity baths. Each thermodynamic neuron requires access to three infinite-capacity baths, associated with the two logical temperatures ($\beta_0$ and $\beta_1$) and with $\beta_m$, as well as one finite-capacity output bath. Since the infinite-capacity baths remain unchanged throughout the computation, they can be shared across all neurons in the network. The number of baths therefore increases with network size only through the finite output baths.

For a dataset with $k$ features, each single-feature network contains six thermodynamic gates. The $k$ single-feature networks therefore require $6k$ gates in total, while the final cascade of $\land$ gates requires a further $k-1$ gates. Since each gate is implemented using $N$ duplicated thermodynamic neurons, the total number of baths is
\begin{equation}
    3+N\bigl[6k+(k-1)\bigr].
    \label{eq:total_gate_number}
\end{equation}
The first term accounts for the three shared infinite-capacity baths, while the second gives the finite-capacity output baths: one for each thermodynamic neuron in the redundant network.

\subsection{Classification performance}
\label{subsec:performance}

We now examine whether our autonomous, stochastic network can learn reliably. Specifically, we compare the classification accuracy of the thermodynamic rule-evaluation engine with that of the standard, deterministic Tsetlin machine, and examine how stochastic gate noise affects performance.
The overall results are summarised in Tab.~\ref{tab:summary}. For all datasets considered, the accuracies of the thermodynamic and standard implementations agree within statistical uncertainty. This shows that reliable classification can be achieved despite the intrinsically stochastic operation of the thermodynamic gates. Importantly, datasets for which the accuracy is lower, such as the tic-tac-toe endgame and adult-income datasets, exhibit this behaviour in both implementations. This indicates that the reduction in performance originates from the representational capacity of the Tsetlin machine itself, rather than from the stochastic dynamics of the thermodynamic realisation (App.~\ref{app:results_discussion}).
Taken together, these results demonstrate that effective learning can be achieved in our autonomous, stochastic architecture without deterministic logical computation, while maintaining comparable classification performance.

\begin{table}[h]
    \centering
    \begin{tabular}{l|cc}
\toprule
 Dataset & Classical & Thermodynamic  \\
\midrule
Mushroom & $\;$ 94.0 ± 2.5 & 93.7 ± 2.6  \\
Breast Cancer $\;$ & $\;$ 91.6 ± 2.3 & 90.3 ± 2.2 \\
Spam & $\;$ 84.2 ± 2.4 & 86.3 ± 2.0 \\
Income & $\;$ 77.2 ± 1.3 & 77.6 ± 2.1 \\
Tic-tac-toe & $\;$ 67.8 ± 3.7 & 68.0 ± 3.0 \\
\bottomrule
\end{tabular}

    \caption{\textbf{Comparison of classification accuracy on unseen data for the standard and thermodynamic Tsetlin machines.} In the thermodynamic implementation, the observation interval is fixed to $T=50$, and each gate is implemented using $N=3$ duplicates. The datasets used, in order of appearance, are Refs.~\cite{mushroom, breast_cancer, spam, tictactoe, adult}.}
    \label{tab:summary}
\end{table}

The results in Tab.~\ref{tab:summary} correspond to thermodynamic gates constructed with $N=3$ duplicates. To understand the role of redundancy in suppressing stochastic gate failures, we examine below how performance varies with $N$ for individual datasets. We focus on the mushroom classification dataset~\cite{mushroom}, in which the task is to distinguish edible from poisonous mushrooms, and the breast cancer diagnostic dataset~\cite{breast_cancer}, in which tumours are classified as benign or malignant.

Performance statistics for the mushroom dataset are shown in Tab.~\ref{tab:mushroom}. In the absence of redundancy ($N=1$), the classification accuracy is substantially reduced, reflecting the high probability of failure when the network is built from single stochastic gates. Increasing the number of duplicates rapidly suppresses these errors: already for $N=2$, the thermodynamic classifier approaches the performance of the standard model, and, for $N\geq 3$, the two agree to within approximately $1\%$.

\begin{table}[htb]
    \centering
    \begin{tabular}{l|cc}
\toprule
 & $\;$ Training Accuracy & Testing Accuracy \\
\midrule
Standard $\;$ & 94.5 ± 2.5 & 94.0 ± 2.5 \\
N=1 & 48.2 ± 0.3 & 48.1 ± 1.3 \\
N=2 & 91.1 ± 2.0 & 90.9 ± 2.6 \\
N=3 & 93.9 ± 2.5 & 93.7 ± 2.6 \\
N=4 & 93.0 ± 3.8 & 92.6 ± 4.3 \\
N=5 & 94.1 ± 3.4 & 94.2 ± 3.3 \\
\bottomrule
\end{tabular}

    \caption{\textbf{Training and testing classification accuracy for the standard and thermodynamic Tsetlin machines on the mushroom classification dataset.} The results are shown as a function of the number of gate duplicates $N$. The dataset contains $8124$ instances, with a class split of $4208$ edible and $3916$ poisonous mushrooms, and $112$ Boolean features in the representation used for training~\cite{mushroom}. The reported accuracies correspond to an 80--20 split between training and test data.}
    \label{tab:mushroom}
\end{table}

A similar pattern is observed for the breast cancer dataset, shown in Tab.~\ref{tab:breast_cancer}. Again, the $N=1$ case exhibits reduced performance, while a small amount of redundancy is sufficient to recover the accuracy of the standard Tsetlin machine. In this case, $N=2$ already achieves this. The quantitative difference with respect to the mushroom dataset reflects differences in dataset complexity, but the qualitative behaviour is the same: modest redundancy is enough to stabilise the stochastic dynamics and restore reliable classification.

\begin{table}[]
    \centering
    \begin{tabular}{l|cc}
\toprule
 & $\;$ Training Accuracy & Testing Accuracy \\
\midrule
Standard $\;$ & 91.7 ± 0.9 & 91.6 ± 2.3 \\
N=1 & 37.8 ± 0.9 & 36.4 ± 4.0 \\
N=2 & 90.5 ± 1.8 & 91.2 ± 3.4 \\
N=3 & 92.1 ± 0.9 & 90.3 ± 2.2 \\
N=4 & 91.6 ± 1.2 & 90.3 ± 3.1 \\
N=5 & 91.1 ± 2.4 & 90.4 ± 2.7 \\
\bottomrule
\end{tabular}

    \caption{\textbf{Training and testing classification accuracy for the standard and thermodynamic Tsetlin machines on the breast cancer diagnostic dataset.} The results are shown as a function of the number of gate duplicates $N$. The original dataset contains $569$ instances, $30$ continuous variables, and a class split of $357$ benign and $212$ malignant tumours~\cite{breast_cancer}. After Booleanisation, each variable is represented by two threshold-based features, yielding \(60\) Boolean features in total. The reported accuracies are obtained using an 80--20 train--test split.}
    \label{tab:breast_cancer}
\end{table}

Redundancy improves not only the mean classification accuracy, but also the statistical stability of the thermodynamic implementation. In particular, for the datasets reported in Tab.~\ref{tab:summary}, the thermodynamic Tsetlin machine, in all but one case (Income dataset), achieves standard deviations that are consistently equal to or smaller than those of the standard model, while maintaining comparable mean accuracy. This indicates that the redundancy built into the thermodynamic gates does more than compensate for stochastic gate failures: it also suppresses performance fluctuations at the network level. In this sense, redundancy acts as an averaging mechanism over stochastic gate outputs, yielding a classifier whose performance is at least as stable as that of its standard counterpart.
The same qualitative behaviour is observed across all datasets included in Tab.~\ref{tab:summary}; additional results are provided in App.~\ref{app:results_discussion}.

Overall, these findings demonstrate that reliable machine learning can emerge from an autonomous thermodynamic network provided that minimal redundancy is introduced to suppress stochastic errors. This establishes that neither external control nor deterministic logic is required for accurate classification.

\section*{Discussion}
\label{sec:discussion}

The results presented in this work show that reliable learning does not require a deterministic computational substrate. Instead, we have shown that an autonomous network of thermodynamic neurons, operating through inherently stochastic transitions, can achieve accuracy comparable to that of the standard Tsetlin machine, provided that appropriate redundancy and thresholding are introduced. In this sense, reliable computation can be realised in a fully physical, noise-driven setting, without requiring deterministic logic or external control.

In our implementation, information is propagated through noisy signal qubits, so logical operations are no longer exact and individual gate errors occur with non-negligible probability. Crucially, however, these errors do not preclude effective learning. In fact, simple architectural features such as thresholding and redundancy are sufficient to suppress stochastic fluctuations at the network level, so that the overall performance is limited not by noise in the substrate, but by the capacity of the learning algorithm itself. This error suppression does, however, come at a cost: achieving high accuracy requires increasing redundancy, leading to a trade-off between reliability and resource efficiency. One possible refinement would be to duplicate only the signal qubits rather than the full gates, so that each neuron would be coupled to $N$ qubits which are monitored as before. This modification would reduce the number of gates by a factor of $N$ [see Eq.~\eqref{eq:total_gate_number}] while retaining the benefits of redundancy, and may offer a more practical route towards large-scale realisation. More broadly, the present model provides a concrete example of how learning can emerge from the same stochastic and dissipative processes that govern physical systems at the microscopic scale. From this perspective, stochasticity is not merely a source of error to be overcome, but a resource that can be shaped and stabilised through network design.

The present analysis is theoretical and focuses on relatively simple network architectures and controlled input distributions. This provides a starting point for exploring how the same thermodynamic framework performs in more complex settings and across a broader range of tasks.
This also points naturally to several directions for future work.

On a fundamental level, the thermodynamic neuron provides a general framework for implementing linearly separable functions \cite{thermodynamic_neuron}. While this work has focused on the logical gates required for classification using the Tsetlin machine ($\lnot$, $\land$, $\lor$), the same principles can be extended to construct more general computational primitives. Combined with the coupling mechanism developed here, this raises the possibility of implementing a much broader class of learning architectures within an autonomous thermodynamic setting.

From a machine-learning perspective, the Tsetlin machine itself admits a wide range of applications, including regression and natural language processing \cite{intro_to_tsetlin,medical_tsetlin,TM-patternrecognition}. An important next step is therefore to investigate whether the stochastic, thermodynamic implementation considered here can support these more complex tasks, and how performance scales with problem difficulty in the presence of noise.

Another promising avenue concerns the role of stochasticity in the learning dynamics themselves. In this work, the feedback applied to the Tsetlin automata is treated as a classical process, while rule evaluation is stochastic. However, recent developments in asymmetric probabilistic Tsetlin machines have shown that introducing stochasticity into the feedback mechanism can improve performance \cite{stochastic_TM}. Extending this idea to the present framework, so that both evaluation and feedback are implemented within a fully stochastic thermodynamic substrate, would represent a natural and potentially powerful generalisation.

It is also important to understand the energetic implications of this approach. The Tsetlin machine is already known for its energy efficiency~\cite{TM-energy}, and embedding it in a thermodynamic framework raises the possibility of analysing explicitly the trade-off between energy consumption and classification accuracy. Such an analysis would provide insight into the thermodynamic efficiency of learning, and could help identify optimal operating regimes that balance performance with physical resource constraints. A complete resource analysis would also include the monitoring apparatus used in the first-passage readout. In particular, the detector records whether an excitation event occurred before the clock tick, and this information must be reset before the device is reused, contributing an additional erasure cost bounded in the ideal limit by $\ln{2}/\beta_\mathrm{reset}$ per erased bit, where $\beta_\mathrm{reset}$ is the inverse temperature of the reservoir used to reset the detector memory. Incorporating these contributions into a full thermodynamic cost model would be a valuable next step towards quantifying the resource requirements of autonomous stochastic learning.

More broadly, the approach developed here is not restricted to Tsetlin machines. Any computation that can be decomposed into networks of linearly separable functions may, in principle, be implemented using thermodynamic neurons. Recent related work has also explored thermodynamic networks as a general framework for physics-based computation using non-equilibrium steady states~\cite{ThermodynamicNetworks}. Together, these perspectives suggest that autonomous, stochastic computation could provide a viable paradigm for machine learning in physical systems where noise and irreversibility are unavoidable.

Within this broader perspective, our focus has been on autonomous stochastic logic in a thermodynamic setting, rather than on quantum computational advantage. The rule-evaluation task considered here is classical and Boolean, while the quantum thermal-machine description provides a microscopic, thermodynamically consistent model for the underlying logic primitives. An important next step is to investigate whether genuinely quantum resources, such as coherence or non-classical correlations, can be incorporated into this framework to provide advantages beyond the classical stochastic setting.

This work shows that accurate learning can arise from an autonomous thermodynamic system governed by stochastic, irreversible dynamics. In this sense, noise and dissipation need not simply limit computation, but can instead play a constructive role in enabling it. This points towards a broader picture in which computation is not imposed on physical systems, but can emerge from their natural dynamics.

\section*{Acknowledgements}
We thank P.~Lipka-Bartosik for useful discussions and S.~Clark and P.~Skrzypczyk for valuable feedback on this manuscript.  \textbf{Funding:} G.R.~acknowledges financial support from the Royal Commission for the Exhibition of 1851 through a Research Fellowship, and from EPSRC through Standard Proposal Grant EP/X016218/1 (Mono-Squeeze). \textbf{Data Availability:} All codes used to produce the data are available at~\cite{Sparrow_ThermoNeuron}.

\filbreak
\renewcommand{\baselinestretch}{1.2}
\bibliography{bibliography}

\setcounter{secnumdepth}{2}
\setcounter{section}{0}
\setcounter{equation}{0}
\setcounter{figure}{0}
\setcounter{table}{0}

\appendix
\setcounter{section}{0}

\renewcommand{\thesection}{\Alph{section}}

\makeatletter
\@addtoreset{equation}{section}
\@addtoreset{figure}{section}
\@addtoreset{table}{section}
\makeatother
\renewcommand{\theequation}{\thesection\arabic{equation}}
\renewcommand{\thefigure}{\thesection\arabic{figure}}
\renewcommand{\thetable}{\thesection\arabic{table}}
\setcounter{secnumdepth}{2}
\setcounter{section}{0}
\setcounter{equation}{0}
\setcounter{figure}{0}
\setcounter{table}{0}

\appendix
\setcounter{section}{0}

\renewcommand{\thesection}{\Alph{section}}

\makeatletter
\@addtoreset{equation}{section}
\@addtoreset{figure}{section}
\@addtoreset{table}{section}
\makeatother
\renewcommand{\theequation}{\thesection\arabic{equation}}
\renewcommand{\thefigure}{\thesection\arabic{figure}}
\renewcommand{\thetable}{\thesection\arabic{table}}

\newcommand{\appsection}[1]{%
  \refstepcounter{section}%
  \section*{APPENDIX \thesection.\ \MakeUppercase{#1}}%
}

\newcommand{\appsubsection}[1]{%
  \refstepcounter{subsection}%
  \subsection*{\thesubsection.\ #1}%
}
\renewcommand{\thesubsection}{\thesection.\arabic{subsection}}

\appendix
\appsection{Gate thermalisation time}
\label{app:thermalisationtime}

The choice of thermalisation time plays an important role in the reliability of the thermodynamic gates. If the observation interval is too short, the gate may fail to register an excitation when required, resulting in a false negative. Conversely, if the interval is too long, unwanted excitations become more likely, leading to false positives instead. An appropriate thermalisation time must therefore balance these competing sources of error.

The cumulative probability of fault-free operation for the single-feature network (Fig.~\ref{fig:network}) is reported in Tab.~\ref{tab:T_network}. This probability is maximised at $T=50$, where $T$ denotes the number of discrete time steps for which the gate is allowed to thermalise.

For a dataset with $n$ features, correct evaluation of the full network requires all single-feature evaluations and all subsequent $\land$ operations to succeed. The probability of fault-free computation for the full network shown in Fig.~\ref{fig:network} is therefore
\begin{equation}
    \mathrm{P(single\text{-}feature\ error\text{-}free)}^n
    \cdot
    \mathrm{P(\land\ error\text{-}free)}^{\,n-1}.
    \label{eq:error_prob}
\end{equation}
This expression provides a lower bound on the probability of a correct final output, since it includes only those realisations in which every gate operates correctly. The final output may indeed remain correct even when some intermediate gates fail, as long as those failures do not alter the overall logical outcome.

Therefore, maximising network performance requires both accurate single-feature evaluation and a high success probability for the $\land$ gates used to combine the outputs of each feature. As shown in Tab.~\ref{tab:T_AND}, the reliability of the $\land$ gate is also maximised at $T=50$.

\begin{table}[htb]
    \centering
    \setlength{\tabcolsep}{4pt}
        \begin{tabular}{l|cccc}
    \toprule
      T & 25 & \textbf{50} & 100 & 150  \\
      \midrule
    P(succ.) & 0.9881 & \textbf{0.9936} & 0.9926 & 0.9917 \\
    P(fail) & 0.0119 & \textbf{0.0064} &	0.0074 & 0.0083\\
    \bottomrule
    \end{tabular}
    \caption{\textbf{Effect of thermalisation time on $\land$-gate success and failure probabilities.} 
    $T$ denotes the number of iterations, each corresponding to a fixed increment in thermalisation time $(10^6)$. The optimal value is highlighted in bold. Probabilities are given to four significant figures.}
    \label{tab:T_AND}
\end{table}

This analysis also illustrates the need for redundancy. Even at the optimal value $T=50$, Eq.~\eqref{eq:error_prob} yields a fault-free probability of only $0.003119$ for $n=60$ for the breast cancer dataset. This value is much lower than the observed classification accuracy because Eq.~\eqref{eq:error_prob} is a conservative lower bound: it counts only those realisations in which every gate operates correctly. Even so, it clearly illustrates how rapidly gate-level errors accumulate as the network size increases, and hence why duplicating gates becomes beneficial.

For completeness, the corresponding dependence on thermalisation time for the $\lnot$ and $\lor$ gates is shown in Tabs.~\ref{tab:T_NOT} and \ref{tab:T_OR}, respectively.

\begin{table}[htb]
    \centering
    \setlength{\tabcolsep}{4pt}
    
\begin{tabular}{l|cccc}
\toprule
      T & 25 & \textbf{50} & 100 & 150  \\
      \midrule
    P(succ.)  & 0.9875 & \textbf{0.9888} & 0.9876 & 0.9864 \\
    P(fail) & 0.0125 & \textbf{0.0112} & 0.0124 & 0.0136\\
    \bottomrule
    \end{tabular}

    \caption{\textbf{Effect of thermalisation time on $\lnot$-gate success and failure probabilities.} Probabilities are given to four significant figures.}
    \label{tab:T_NOT}
\end{table}

\begin{table}[htb]
    \centering
    \setlength{\tabcolsep}{4pt}
        \begin{tabular}{l|cccc}
    \toprule
      T & \textbf{25} & 50 & 100 & 150  \\
      \midrule
    P(succ.) & \textbf{0.9824} &	0.9818 & 0.9794 & 0.9770 \\
    P(fail) & \textbf{0.0176} & 0.0182 & 0.0206 &	0.0230\\
    \bottomrule
    \end{tabular}
    \caption{\textbf{Effect of thermalisation time on $\lor$-gate success and failure probabilities.} Probabilities are given to four significant figures.}
    \label{tab:T_OR}
\end{table}

\appsection{Dataset-specific performance}
\label{app:results_discussion}

As discussed in Sec.~\ref{subsec:performance}, the classification accuracy of the stochastic implementation depends strongly on the amount of redundancy introduced at the gate level. Without duplication, performance is substantially reduced, whereas with only modest redundancy the accuracy of the stochastic implementation becomes statistically indistinguishable from that of the standard Tsetlin machine. This qualitative behaviour is observed across all datasets considered here, including those for which the baseline accuracy is itself relatively modest. In this Appendix, we comment briefly on the dataset-specific trends underlying these results.

\subsection{Spam dataset}

For the spam dataset (Table~\ref{tab:spam}), the standard Tsetlin machine achieves an accuracy above $80\%$. Reproducing this level of performance in the stochastic implementation requires $N \geq 3$, while the cases $N=1$ and $N=2$ lead to noticeably lower accuracy. This behaviour is consistent with the trends observed for the mushroom and breast-cancer datasets, and again reflects the role of redundancy in suppressing stochastic gate errors.

The dataset consists of continuous features, which must be discretised before being processed by the Tsetlin machine. In the present analysis, each continuous feature is represented using four threshold conditions, namely whether the feature value is less than or equal to $x_1$, $x_2$, $x_3$, and $x_4$. This discretisation yields an accuracy above $80\%$. Although a finer partitioning might improve performance, it would also increase the effective complexity of the Boolean representation and could therefore affect both model capacity and the risk of overfitting. This was not explored further here, since our focus is on the effect of stochasticity in the thermodynamic implementation.

\begin{table}[htb]
    \centering
    \begin{tabular}{l|cc}
\toprule
 & $\;$ Training Accuracy & Testing Accuracy \\
\midrule
Standard $\;$  & 84.5 ± 1.8 & 84.2 ± 2.4 \\
N=1 & 39.4 ± 0.3 & 39.3 ± 1.3 \\
N=2 & 61.8 ± 4.3 & 63.3 ± 4.9 \\
N=3 & 86.1 ± 1.7 & 86.3 ± 2.0 \\
N=4 & 85.6 ± 1.8 & 85.4 ± 3.0 \\
N=5 & 85.6 ± 2.2 & 85.5 ± 2.1 \\
\bottomrule
\end{tabular}

    \caption{\textbf{Training and testing classification accuracy for the standard and thermodynamic Tsetlin machines on the spam dataset.} The results are shown as a function of the number of gate duplicates $N$. The dataset contains $4601$ instances, with a class split of $1813$ spam and $2788$ non-spam emails, and $228$ Boolean features obtained using four intervals per continuous feature~\cite{spam}. The reported accuracies are obtained using an 80--20 train--test split.}
    \label{tab:spam}
\end{table}

\subsection{Tic-tac-toe dataset}

For the tic-tac-toe dataset (Table~\ref{tab:tictactoe}), the overall accuracy is lower and exhibits a larger variance than for the other datasets considered. The close agreement between training and testing performance suggests that the limitation arises from the capacity of the model rather than from overfitting. Since the same behaviour is observed for both the standard and thermodynamic versions, this indicates that the reduced accuracy should be attributed primarily to the Tsetlin-machine model on this dataset, rather than to the stochastic dynamics of the thermodynamic network.

Despite the lower absolute accuracy, the dependence on redundancy follows the same qualitative pattern as elsewhere: performance is reduced for \(N=1\), but for \(N \geq 2\) the thermodynamic implementation matches the standard result within statistical uncertainty. This again supports the conclusion that stochasticity does not impose an additional limitation beyond that already present in the underlying learning architecture.

\begin{table}[htb]
    \centering
    \begin{tabular}{l|cc}
\toprule
 & $\;$ Training Accuracy & Testing Accuracy \\
\midrule
Standard $\;$ & 68.3 ± 2.4 & 67.8 ± 3.7 \\
N=1 & 59.2 ± 2.4 & 57.9 ± 2.4 \\
N=2 & 66.2 ± 1.3 & 67.4 ± 3.4 \\
N=3 & 66.7 ± 2.4 & 68.0 ± 3.0 \\
N=4 & 67.0 ± 3.4 & 67.2 ± 5.6 \\
N=5 & 68.4 ± 2.7 & 69.1 ± 2.7 \\
\bottomrule
\end{tabular}

    \caption{\textbf{Training and testing classification accuracy for the standard and thermodynamic Tsetlin machines on the tic-tac-toe dataset.} The results are shown as a function of the number of gate duplicates $N$. The dataset contains $958$ instances, with a class split of $332$ negative and $626$ positive outcomes, and $27$ Boolean features~\cite{tictactoe}. The reported accuracies are obtained using an 80--20 train--test split.}
    \label{tab:tictactoe}
\end{table}

\subsection{Income dataset}

A similar picture emerges for the income dataset (Table~\ref{tab:income}), for which the overall accuracy is again relatively low in both implementations. As in the tic-tac-toe case, this points to a limitation of the Tsetlin-machine model on this task rather than of the stochastic thermodynamic realisation itself.

The income dataset contains both continuous and categorical features and therefore requires booleanisation before training. Although the precise choice of discretisation and encoding may affect performance, this is not the focus of the present analysis. The relevant observation here is that the reduced accuracy is already present in the standard implementation, and therefore cannot be attributed to thermodynamic noise. As before, redundancy restores the performance of the stochastic implementation to the level of the standard Tsetlin machine.

\begin{table}[htb]
    \centering
    \begin{tabular}{l|cc}
\toprule
 & $\;$ Training Accuracy & Testing Accuracy \\
\midrule
Standard  $\;$& 77.2 ± 1.2 & 77.2 ± 1.3 \\
N=1 & 23.9 ± 0.1 & 23.9 ± 0.3 \\
N=2 & 76.2 ± 0.6 & 76.2 ± 0.8 \\
N=3 & 77.6 ± 2.0 & 77.6 ± 2.1 \\
N=4 & 77.5 ± 1.6 & 77.4 ± 2.0 \\
N=5 & 77.9 ± 1.8 & 77.8 ± 1.9 \\
\bottomrule
\end{tabular}

    \caption{\textbf{Training and testing classification accuracy for the standard and thermodynamic Tsetlin machines on the income dataset.} The results are shown as a function of the number of gate duplicates $N$. The dataset contains $48842$ instances, with a class split of $37155$ below \$50K and $11687$ above \$50K, and $126$ Boolean features obtained using four intervals per continuous feature~\cite{adult}. The reported accuracies are obtained using an 80--20 train--test split.
    }
    \label{tab:income}
\end{table}

\clearpage
\onecolumngrid
\setcounter{section}{0}
\setcounter{equation}{0}
\setcounter{figure}{0}
\setcounter{table}{0}

{\centering \Large\bfseries Supplementary Materials\par}

\renewcommand{\thesection}{S}

\renewcommand{\theequation}{S\arabic{equation}}
\renewcommand{\thefigure}{S\arabic{figure}}
\renewcommand{\thetable}{S\arabic{table}}

\setcounter{secnumdepth}{2}
\setcounter{section}{0}
\setcounter{subsection}{0}
\setcounter{equation}{0}
\setcounter{figure}{0}
\setcounter{table}{0}

\renewcommand{\thesection}{S\arabic{section}}
\renewcommand{\thesubsection}{S\arabic{section}.\arabic{subsection}}

\makeatletter
\@addtoreset{equation}{section}
\@addtoreset{figure}{section}
\@addtoreset{table}{section}
\makeatother

\renewcommand{\theequation}{S\arabic{equation}}
\renewcommand{\thefigure}{S\arabic{figure}}
\renewcommand{\thetable}{S\arabic{table}}

\newcommand{\SMsection}[1]{%
  \refstepcounter{section}%
  \vspace{2.5em}%
  {\centering \bfseries S\arabic{section}. #1\par}%
  \vspace{1.5em}%
}

\newcommand{\SMsubsection}[1]{%
  \refstepcounter{subsection}%
  \vspace{2.5em}
  \subsection*{\thesubsection\ #1}%
  \nobreak\addvspace{1.5\baselineskip}
}

\SMsection{Parametrisation of the steady-state output}
\label{SM:steady_state_out}

In this Supplementary Note, we provide additional algebraic details for the steady-state output discussed in Sec.~\ref{subsec:neuron}. In particular, we show how imposing the logical output values $\beta_{\min}$ and $\beta_{\max}$ determines the parameters $\Delta$ and $\beta_m$ entering the steady-state expression for $\beta_z^\infty$.

We start from Eq.~\eqref{eq:excited_output}. Introducing the parameter
\begin{equation}
    \Delta := \frac{\mu}{\mu+\mu'},
\end{equation}
(which will be shown below to coincide with the definition used in the main text) the steady-state population can be written as
\begin{equation}
    g_z(\beta_z^\infty)=\Delta\, g_z(\beta_v)+(1-\Delta)\, g_z(\beta_m).
\label{eq:excited_pop_output}
\end{equation}
To ensure logical behaviour, we impose the boundary conditions of Eq.~\eqref{eq:MaxMin_definition}, so that the two extreme limits of the virtual temperature map to the logical output values $\beta_{\min}$ and $\beta_{\max}$. Eq.~\eqref{eq:excited_pop_output} then gives
\begin{subequations}
\begin{align}
    g_z(\beta_{\min}) &= \Delta + (1-\Delta)\, g_z(\beta_m), \\
    g_z(\beta_{\max}) &= (1-\Delta)\, g_z(\beta_m). \label{eq:g_z-beta_max}
\end{align}
\end{subequations}
Subtracting the two equations yields
\begin{equation}
    \Delta = g_z(\beta_{\min})-g_z(\beta_{\max}) \geq 0.
\label{eq:Delta_definition_appendix}
\end{equation}
From Eq.~\eqref{eq:g_z-beta_max} and using the explicit form of $g_z(x)$, we obtain
\begin{equation}
    \beta_m=\frac{1}{\epsilon_z}\mathrm{log}\bigl[(1-\Delta)e^{\beta_{\max}\epsilon_z}-\Delta\bigr].
\end{equation}
Using Eq.~\eqref{eq:g_z-beta_max}, the steady-state population can be written as in Eq.~\eqref{eq:g_z-beta_z-infty}. Inverting the function $g_z(\beta_z^\infty)$ then gives Eq.~\eqref{eq:beta_z^inf}.

\SMsection{Virtual temperatures for $\land$ and $\lor$ gates}
\label{SM:virtual_temp}

In this section, we derive the virtual temperatures used to implement the $\land$ and $\lor$ gates in the thermodynamic network. The construction follows the general procedure introduced in Ref.~\cite{thermodynamic_neuron}, in which a linearly separable Boolean function is associated with a separating hyperplane, and the corresponding weight vector determines the virtual temperature of the gate.

For a Boolean function of $n$ inputs, let $D$ denote the corresponding set of input-output tuples $(\beta_1,\ldots,\beta_n,\beta_z)$, where $\beta_z\in\{0,1\}$ is the logical output. If the dataset is linearly separable, one may introduce a separating hyperplane with weight vector $\mathbf{w}=(w_0,\ldots,w_n)$. In the thermodynamic-neuron construction, the virtual temperature is then chosen to reproduce the corresponding affine function of the inputs. Following Ref.~\cite{thermodynamic_neuron}, this is written as
\begin{equation}
    \beta_v = \frac{\alpha}{\epsilon_z}\left[w_0 + \sum_{k=1}^{n} w_k \beta_k\right],
    \label{eq:virtual_temp_definition}
\end{equation}
where $\alpha$ is a scaling parameter that sets the overall energy scale. As discussed in Supplementary Note~\ref{app:alpha}, increasing $\alpha$ sharpens the gate response and improves the approximation to an ideal logical element.

\subsection{AND}

We first consider the $\land$ gate. The corresponding input-output dataset is
\begin{equation}
    D := \bigl\{(0,0,0), (0,1,0), (1,0,0), (1,1,1)\bigr\}.
\end{equation}
This dataset is linearly separable. A convenient separating hyperplane for this dataset is $\frac{2}{3}\,(\beta_1+\beta_2)=1$, since the point $(1,1)$ lies on the $1$-output side, while $(0,0)$, $(0,1)$, and $(1,0)$ lie on the $0$-output side. The output $1$ is then assigned when $\frac{2}{3}\,(\beta_1+\beta_2)>1$, as illustrated in Fig.~\ref{fig:AND_gate}. The corresponding weight vector is therefore $\mathbf{w}=\bigl(-1,\frac{2}{3},\frac{2}{3}\bigr)$. Substituting these weights into Eq.~\eqref{eq:virtual_temp_definition} yields the virtual temperature for the $\land$ gate,
\begin{equation}
    \beta_v = \frac{\alpha}{\epsilon_z}\left(-1 + \frac{2}{3}\beta_1 + \frac{2}{3}\beta_2\right).
    \label{eq:virtual_temp_AND}
\end{equation}

\subsection{OR}
\label{app:virtual_temp_OR}

\begin{figure}[htb]
    \centering
    \begin{subfigure}[t]{0.48\linewidth}
        \centering
        \includegraphics[width=\linewidth]{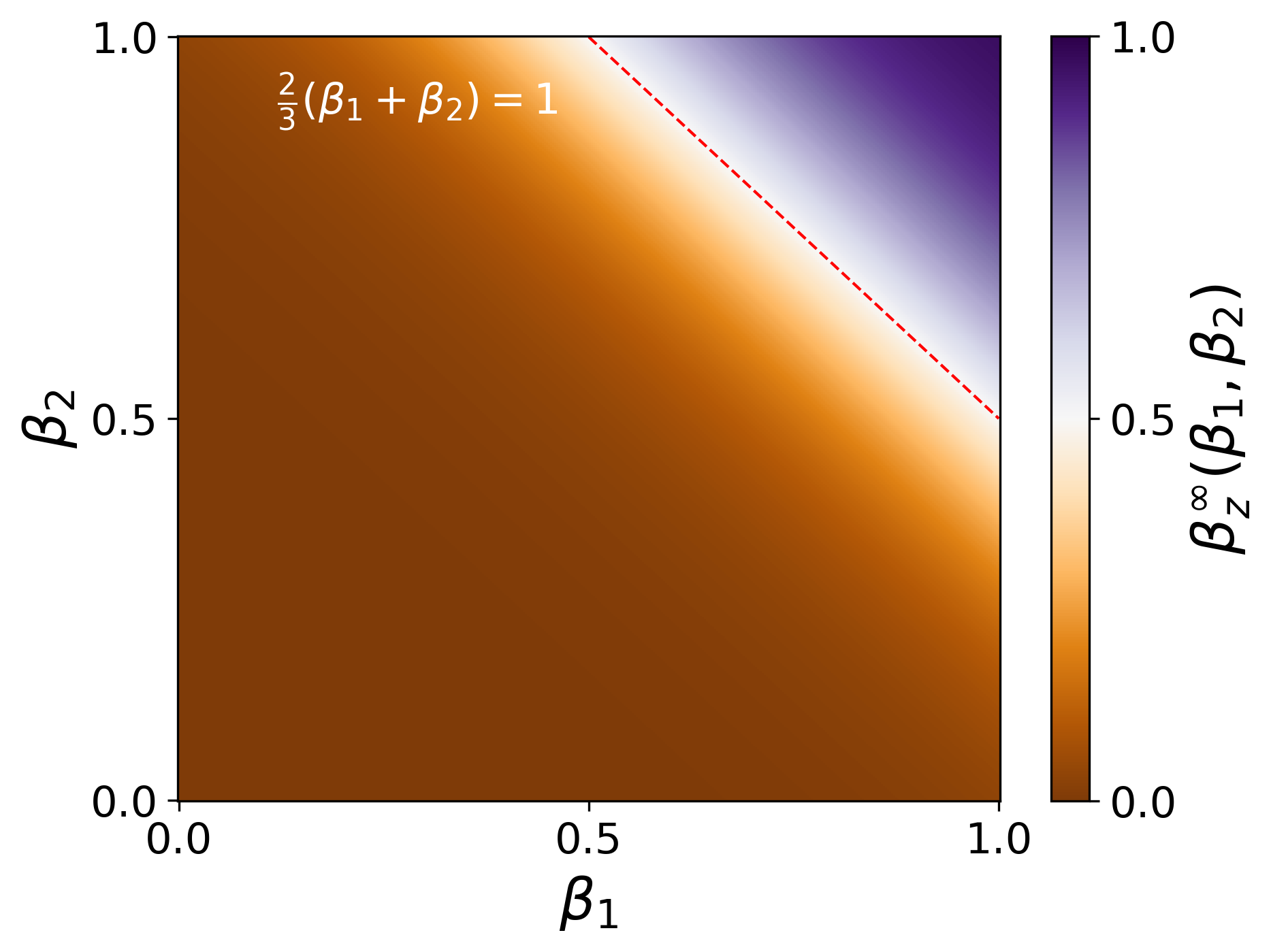}
        \caption{AND}
        \label{fig:AND_gate}
    \end{subfigure}
    \hfill
    \begin{subfigure}[t]{0.48\linewidth}
        \centering
        \includegraphics[width=\linewidth]{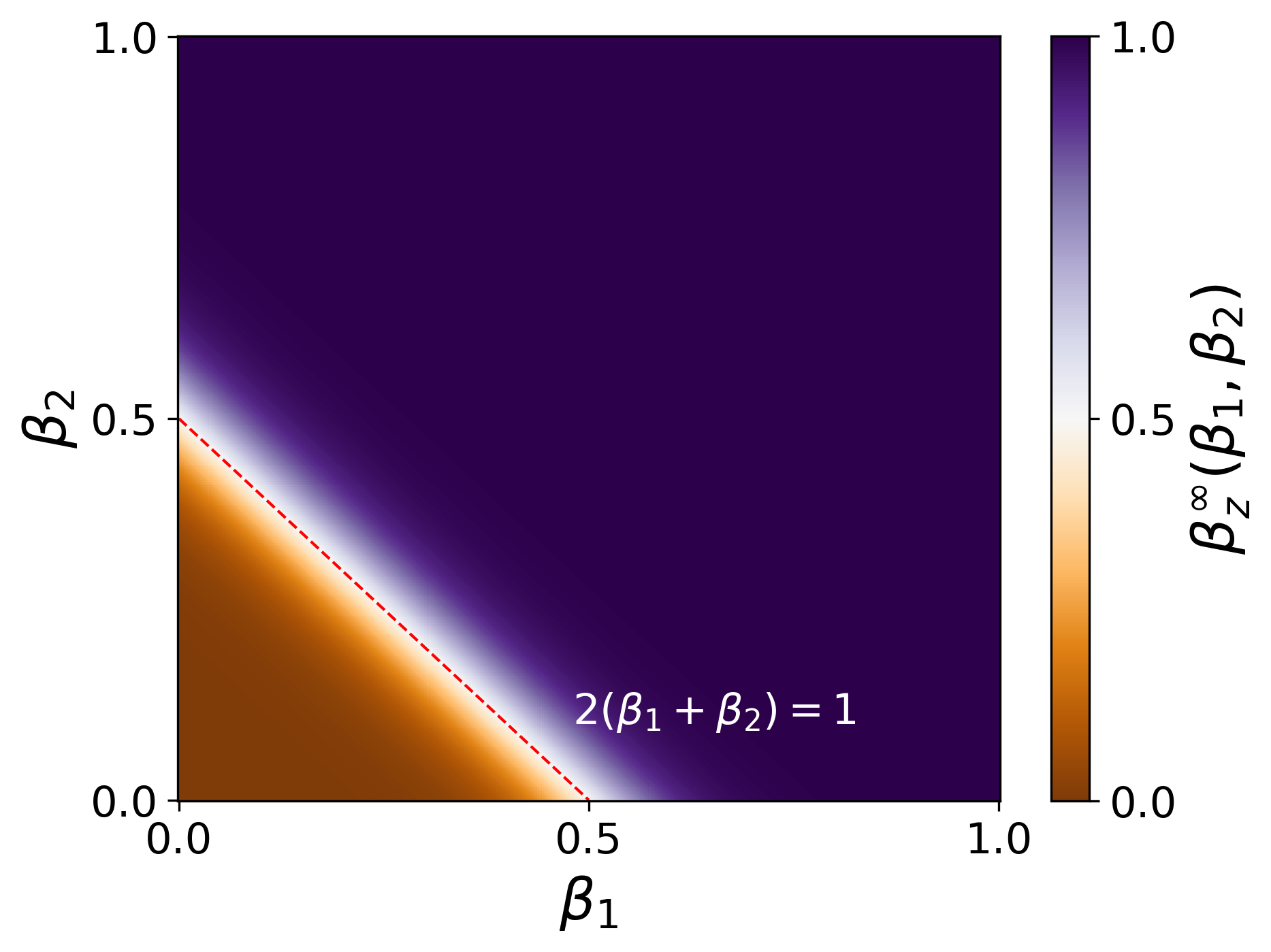}
        \caption{OR}
        \label{fig:OR_gate}
    \end{subfigure}
    \caption{\textbf{Steady-state output temperature for the $\land$ and $\lor$ gates.} The quantity $\beta_z^{\infty}$, given by Eq.~\eqref{eq:beta_z^inf}, is shown as a function of the inputs for the virtual temperatures defined in Eqs.~\eqref{eq:virtual_temp_AND} and \eqref{eq:virtual_temp_OR}. Parameters: $\epsilon_z = 0.1$, $\alpha = 10$, $\epsilon_0 = \alpha(\epsilon_z+4)$, $\epsilon_1 = 2\alpha$, $\epsilon_2 = 2\alpha$, $\beta_0 = 1/(\epsilon_z+4)$, $\beta_\text{min} = 0$, and $\beta_\text{max} = 1$.}
    \label{fig:combined}
\end{figure}

We now consider the $\lor$ gate, for which the corresponding dataset is
\begin{equation}
    D := \bigl\{(0,0,0), (0,1,1), (1,0,1), (1,1,1)\bigr\}.
\end{equation}
Again, the dataset is linearly separable. In this case, a suitable separating hyperplane is $2\beta_1 + 2\beta_2 = 1$, with output $1$ assigned when $2\beta_1 + 2\beta_2 > 1$, as shown in Fig.~\ref{fig:OR_gate}. This gives the weight vector $\mathbf{w}=\bigl(-1,2,2\bigr)$. The corresponding virtual temperature is therefore
\begin{equation}
    \beta_v = \frac{\alpha}{\epsilon_z}\bigl(-1 + 2\beta_1 + 2\beta_2\bigr).
    \label{eq:virtual_temp_OR}
\end{equation}
Substituting this into Eq.~\eqref{eq:beta_z^inf} gives the steady-state output behaviour shown in Fig.~\ref{fig:OR_gate}.

\SMsection{Effect of \texorpdfstring{$\alpha$}{alpha} on gate quality}
\label{app:alpha}

\begin{figure}[htb]
    \centering
    \includegraphics[width=.5\linewidth]{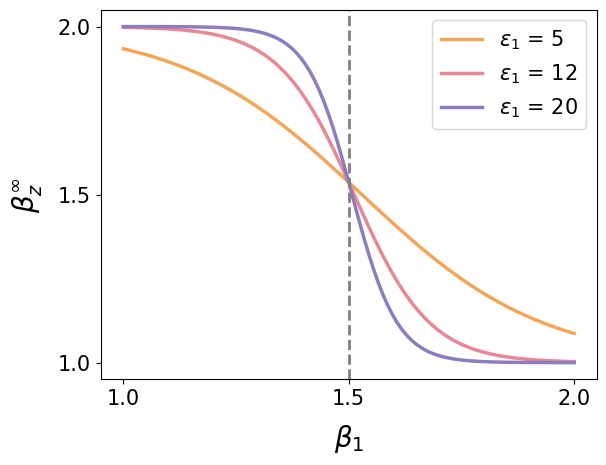}
    \caption{\textbf{Effect of varying $\epsilon_1$ for the $\lnot$-gate.} As $\epsilon_1$ increases, the steady-state output approaches the ideal sigmoid response. Parameters: $\epsilon_z=0.1$, $\beta_0 = 1.5$, $\beta_\text{min} = 1$, and $\beta_\text{max} = 2$.}
    \label{fig:vary epsilon1}
\end{figure}

\begin{figure}[ht]
    \centering
    \begin{subfigure}[t]{0.48\linewidth}
        \centering
        \includegraphics[width=\linewidth]{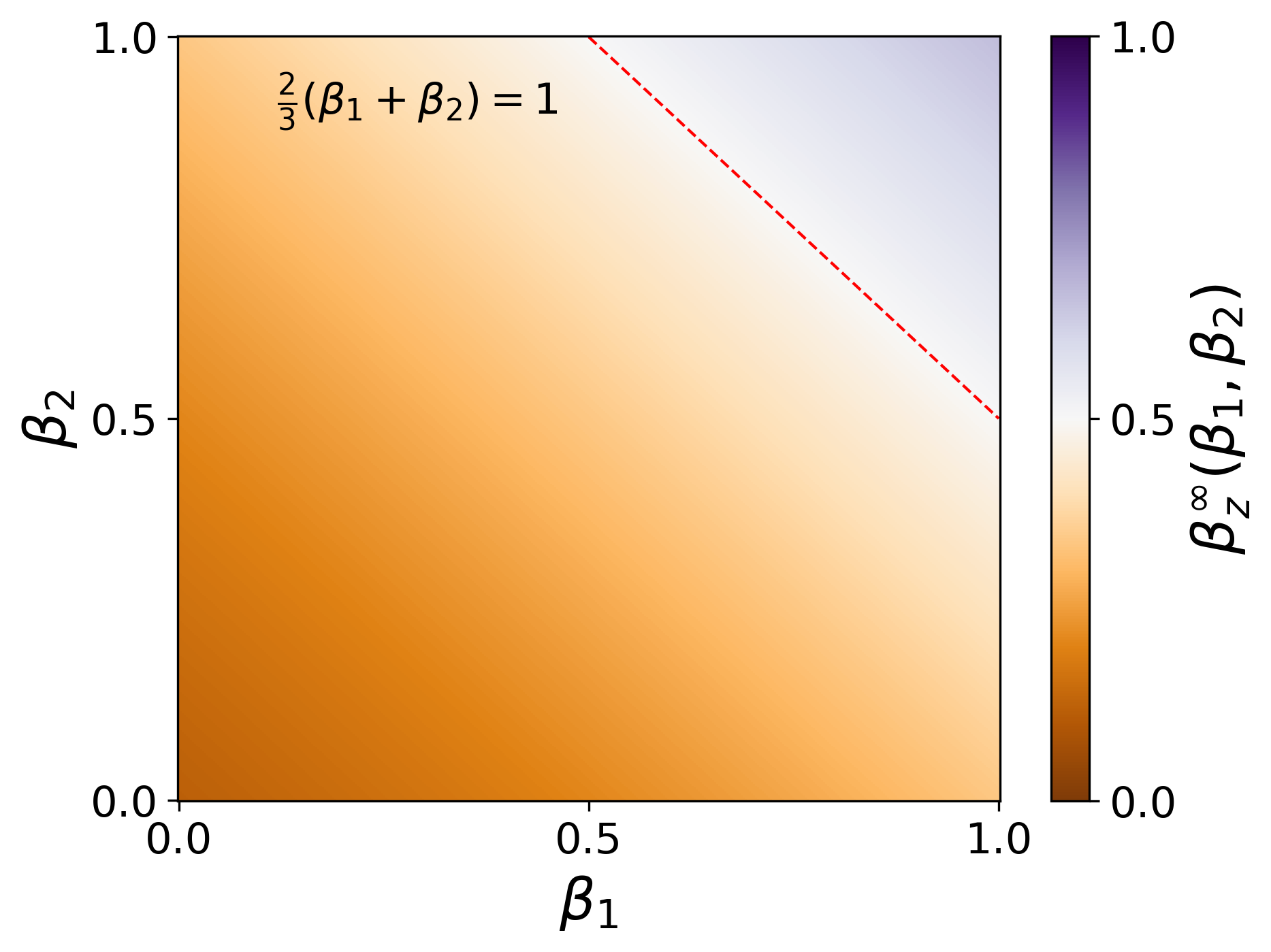}
        \caption{$\alpha = 2$}
        \label{fig:AND_alpha2}
    \end{subfigure}
    \hfill
    \begin{subfigure}[t]{0.48\linewidth}
        \centering
        \includegraphics[width=\linewidth]{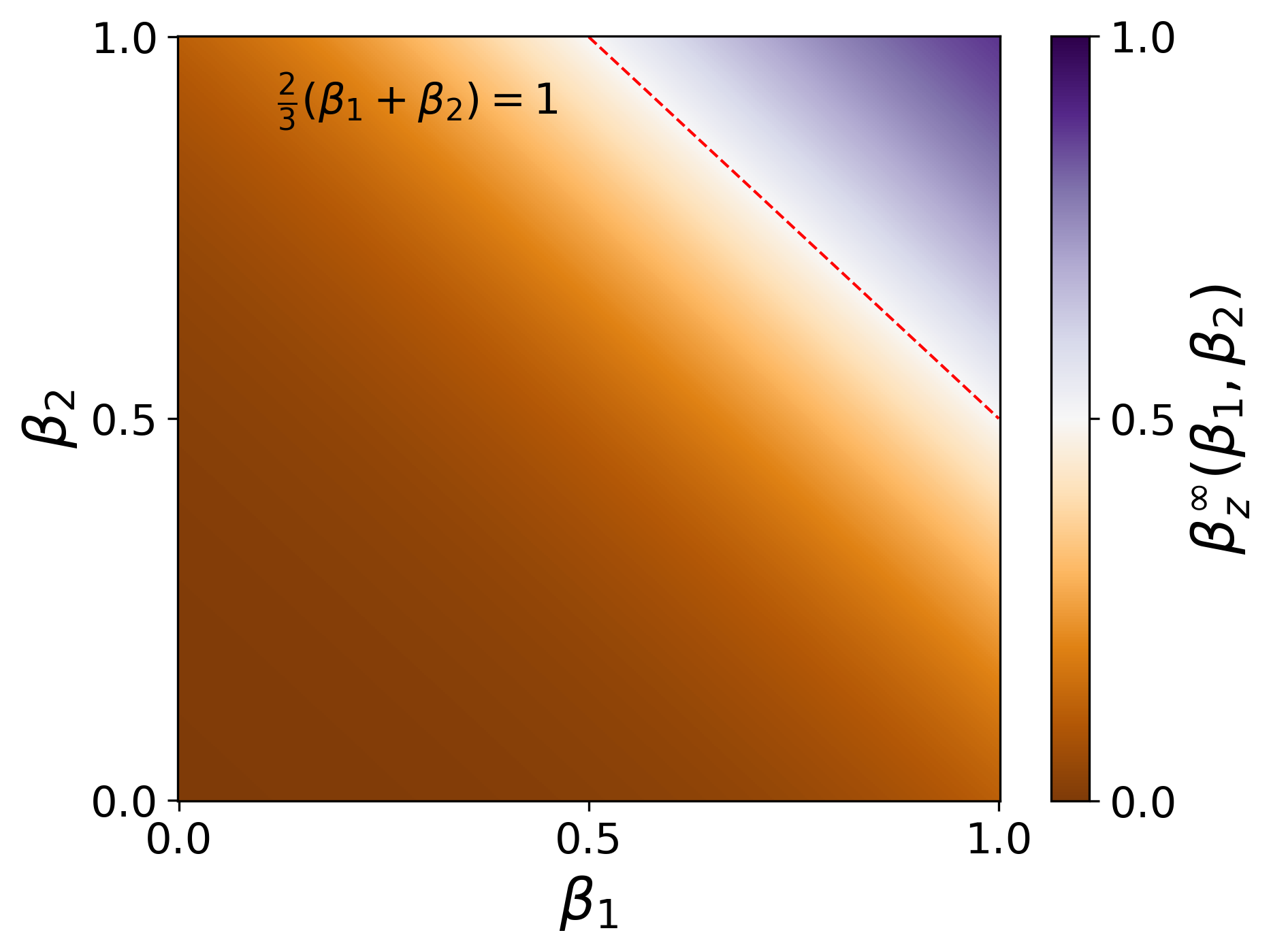}
        \caption{$\alpha = 6$}
        \label{fig:AND_alpha6}
    \end{subfigure}
    \caption{\textbf{Effect of varying $\alpha$ for the $\land$-gate.} Increasing $\alpha$ sharpens the transition across the separating hyperplane and drives the output more closely towards its logical limiting values. Parameters: $\epsilon_z = 0.1$, $\boldsymbol{\epsilon}=\alpha(\epsilon_z +4, 2, 2)$, $\beta_0 = 1/(\epsilon_z+4)$, $\beta_\text{min}=0$, and $\beta_\text{max}=1$.}
    \label{fig:increase_alpha}
\end{figure}

This section examines how the energy-scale parameters of the thermodynamic gates affect the sharpness of their logical response, and hence the quality of the approximation to an ideal Boolean element. For the $\lnot$-gate, increasing $\epsilon_1$ sharpens the transition between the two logical output values, so that the steady-state response approaches the ideal sigmoid more closely, as shown in Fig.~\ref{fig:vary epsilon1}. For the thermodynamic $\land$ and $\lor$ gates, the parameter $\alpha$ plays an analogous role. This is illustrated for the $\land$ gate in Fig.~\ref{fig:increase_alpha}. Since $\alpha$ sets the overall energy scale, it rescales the virtual temperature and therefore controls the effective sharpness of the gate response. As $\alpha$ increases, the transition between logical outputs becomes steeper, the output values move closer to their ideal logical limits, and the intermediate region around the separating hyperplane becomes narrower (see also Fig.~\ref{fig:AND_gate} for comparison). In the colour plots, this corresponds to a more saturated separation between the two logical regions and a thinner crossover band.
Thus, increasing $\alpha$ improves gate quality in much the same way as increasing $\epsilon_1$ does for the $\lnot$-gate: in both cases, the thermodynamic response approaches the behaviour of an ideal logical element more closely. At the same time, this improvement comes at the cost of increasing the underlying energy scale, thereby highlighting a trade-off between gate quality and energetic cost.

\FloatBarrier

\end{document}